\documentclass[12pt]{article}
\usepackage{amsmath,amssymb,bm}
\usepackage{graphicx}
\usepackage{enumerate}
\usepackage{natbib}
\usepackage[colorlinks=true,allcolors=black]{hyperref}%
\usepackage{url} 
\usepackage[percent]{overpic}
\usepackage{caption}
\usepackage{graphicx,subfigure}
\usepackage{dsfont}
\usepackage{algorithm,algorithmic}
\usepackage[normalem]{ulem}
\usepackage{xcolor}
\usepackage{epstopdf,multirow,lscape,comment}

\newcommand{\blind}{0}

\newtheorem{Th}{{\bf Theorem}}

\newtheorem{Cor}{{\bf Corollary}}

\newtheorem{Ass}{Assumption}

\def\ba{{\mathbf a}}
\def\bb{{\mathbf b}}
\def\bB{{\mathbf B}}

\def\bD{{\mathbf D}}
\def\bE{{\mathbf E}}
\def\b1e{{\mathbf e}}

\def\bI{{\mathbf I}}

\def\bB{{\mathbf B}}
\def\bM{{\mathbf M}}

\def\bQ{{\mathbf Q}}

\def\bR{{\mathbf R}}
\def\br{{\mathbf r}}
\def\bS{{\mathbf S}}

\def\by{{\mathbf y}}

\def\bz{{\mathbf z}}
\def\m{{\mathbf m}}
\def\bx{{\mathbf x}}
\def\bX{{\mathbf X}}

\def\bv{{\mathbf v}}
\def\bV{{\mathbf V}}

\def\bW{{\mathbf W}}

\def\bzero{{\mathbf 0}}

\def\bLambda{{\boldsymbol{\Lambda}}}

\def\bbeta{{\boldsymbol{\beta}}}

\def\bTheta{{\boldsymbol{\Theta}}}

\def\boxit#1{\vbox{\hrule\hbox{\vrule\kern6pt
          \vbox{\kern6pt#1\kern6pt}\kern6pt\vrule}\hrule}}

\def\wt{\widetilde}

\def\wh{\widehat}

\def\bse{\begin{eqnarray*}}
\def\ese{\end{eqnarray*}}
\def\be{\begin{eqnarray}}
\def\ee{\end{eqnarray}}
\def\bq{\begin{equation}}
\def\eq{\end{equation}}
\def\bse{\begin{eqnarray*}}
\def\ese{\end{eqnarray*}}

\def\wh{\widehat}

\def\b1e{{\mathbf e}}

\def\bx{{\mathbf x}}
\def\bX{{\mathbf X}}

\def\bS{{\mathbf S}}

\def\bzero{{\mathbf 0}}

\newcommand*{\mydot}{\mathrel{\scalebox{0.4}{$\bullet$}}}

\newcommand{\bSigma}{\mbox{\boldmath $\Sigma$}}

\def\rkcomment#1{\vskip 2mm\boxit{\vskip 2mm{\color{blue}\bf#1} {\color{blue}\bf -- RK\vskip 2mm}}\vskip 2mm}
\newcommand{\rkc}[1]{{\color{red}[RK: #1]}}

\def\rkcomment#1{\vskip 2mm\boxit{\vskip 2mm{\color{blue}\bf#1} {\color{blue}\bf -- RK\vskip 2mm}}\vskip 2mm}
\newcommand{\rkcr}[1]{{\color{black}#1}}

\begin{document}

\if0\blind
{
  \title{\bf High-dimensional covariance regression with application to co-expression QTL detection}
  \author{Rakheon Kim\\
    Department of Statistical Science, Baylor University\\
    and \\
    Jingfei Zhang \\
    Goizueta Business School, Emory University}
    \date{}
  \maketitle
} \fi

\if1\blind
{
  \bigskip
  \bigskip
  \bigskip
  \begin{center}
    {\LARGE\bf Title}
\end{center}
  \medskip
} \fi

\bigskip
\begin{abstract}
\baselineskip=17.5pt
While covariance matrices have been widely studied in many scientific fields, relatively limited progress has been made on estimating conditional covariances that permits a large covariance matrix to vary with high-dimensional subject-level covariates. In this paper, we present a new sparse covariance regression framework that models the covariance matrix as a function of subject-level covariates. In the context of co-expression quantitative trait locus (QTL) studies, our method can be used to determine if and how gene co-expressions vary with genetic variations. To accommodate high-dimensional responses and covariates, we stipulate a combined sparsity structure that encourages covariates with non-zero effects and edges that are modulated by these covariates to be simultaneously sparse. We approach parameter estimation with a blockwise coordinate descent algorithm, and investigate the $\ell_1$ and $\ell_2$ convergence rate of the estimated parameters. In addition, we propose a computationally efficient debiased inference procedure for uncertainty quantification. The efficacy of the proposed method is demonstrated through numerical experiments and an application to a gene co-expression network study with brain cancer patients.

\end{abstract}

\noindent%
\vfill

\newpage
\baselineskip=26.5pt

\section{Introduction}
\label{sec:intro}
A covariance matrix measures the associations amongst a set of variables and its estimation and analysis play an important role in a wide range of applications, 
such as genetics \citep{butte2000discovering,su2023cell}, neuroscience \citep{zhang2020mixed,zhang2023generalized}, finance \citep{el2010high, xue2012positive} and climatology \citep{bickel2008covariance}. 
For example, in genetics, the covariance matrix estimated from gene expressions across different biological samples, often referred to as a co-expression network, is routinely used in identifying functional gene modules and dysregulated pathways in disease \citep{langfelder2008wgcna,su2023cell}.
Although most co-expression analyses to date assume a common covariance matrix for different subjects, the structure and degree of covariance may depend on individual's characteristics such as age, sex and genotype, which are referred to as individual-level covariates or covariates in this paper when there is no ambiguity. For example, it is known that co-expressions among genes can be affected by individual genetic variants, clinical and environmental factors \citep{van2018integrative}. In particular, a genetic variant that affects co-expressions between a pair of genes is termed a co-expression quantitative trait loci (QTL). Identifying co-expression QTLs is of great scientific interests and can be crucial in developing gene therapies that target specific gene or pathway disruptions \citep{van2018integrative,zhang2023eqtl}. 

Although the literature on estimating large covariance matrices is steadily increasing \citep[][and others]{wu2003nonparametric, huang2006covariance,bickel2008covariance, bickel2008regularized, rothman2009generalized, lam2009sparsistency, bien2011sparse}, the majority of existing methods assume a homogeneous population obeying a common covariance model.
Some others have considered modeling covariate-dependent covariance matrices.
For example, \citet{anderson1973asymptotically} modeled the covariance matrix $\bSigma\in\mathbb{R}^{p\times p}$ as a linear combination of a given set of symmetric matrices; \citet{chiu1996matrix} modeled elements in the logarithm of $\bSigma$, denoted as $\log\bSigma$, as a linear function of covariates $\bx\in\mathbb{R}^{q}$. 
As noted by the authors, parameter interpretation for this model can be difficult, as a submatrix of $\bSigma$ is not generally the matrix exponential of the same submatrix of $\log\bSigma$, and so the entries in $\log\bSigma$ do not directly relate to the corresponding entries in $\bSigma$. 
\citet{pourahmadi1999joint} modeled elements of the Cholesky decomposition of $\bSigma^{-1}$ as linear functions of $\bx$, though this model is not invariant to the reorderings of response variables. 
\citet{zou2017covariance} related $\bSigma$ to a linear combination of similarity matrices of covariates. However, the covariates considered in this work are variable-specific and not individual-specific (e.g., covariates of genes but not of individuals). As such, the estimated covariance could not account for individual-level heterogeneity due to clinical covariates and genotypes. 
\rkcr{More recent work by \citet{zou2022inference} extends this method to account for individual-level heterogeneity by allowing the similarity matrices to vary across individuals. However, both \citet{zou2017covariance} and \citet{zou2022inference} assume that the similarity matrices are known, which may not be available in our motivating data example.}

Notably, \citet{hoff2012covariance} proposed to model $\bSigma$ as a quadratic function of covariates $\bx$ written as $\bB\bx\bx^\top\bB^\top$, $\bB\in\mathbb{R}^{p\times q}$, which also admits a nice random-effects model representation; model estimation is carried out using the expectation–maximization (EM) algorithm or an Markov chain Monte Carlo (MCMC) via Gibbs sampling. Their modeling framework is further extended in \citet{fox2015bayesian} by considering non-linear effects, in \citet{franks2021reducing} by considering high dimensional response variables and in \citet{alakus2022covariance} by considering random forests. 
The computational costs of the above extended methods can be prohibitive when dimensions of the response variables and covariates are both high. Moreover, due to the quadratic form of $\bB\bx\bx^\top\bB^\top$, sparsity in parameter $\bB$ does not directly translate to sparse effects of covariates, possibly limiting model interpretability. 
\citet{zhao2021covariate,park2023bayesian} studied a principal regression approach that models $\bm\gamma^\top\bSigma\bm\gamma$, where $\bm\gamma$ is an unknown rotation vector, as a generalized linear model of $\bx$. Parameter interpretation for this model may not be straightforward, as elements in $\bSigma$ are not directly modeled as a function of $\bx$. It is also challenging to further extend this approach to the high dimensional setting. 


To flexibly model large covariance matrices modulated by individual-level covariates, we propose a covariance regression model that allows the structure and degree of covariance to vary with discrete and continuous covariates of high dimensions. Specifically, the covariance matrix is modeled as a linear function of covariates with matrix-valued coefficients, subject to constraints that ensure positive semi-definiteness. Our model needs not to make specific assumptions on the distribution of response variables, such as the Gaussian assumption imposed in \citet{hoff2012covariance}. Using method of moments, we formulate coefficient estimation as a least squares problem and impose a sparse group lasso penalty that simultaneously encourages effective covariates and their effects on the covariance matrix to be sparse. This combined sparsity assumption facilitates model estimability and interpretability, and is closely connected with multi-tasking learning \citep{argyriou2008convex}. 
In theory, we investigate the convergence rate of the proposed estimator, allowing both the response variables and covariates to be high-dimensional. 
\rkcr{Our theoretical analysis involves both the variance and covariance terms, treated differently in the penalty function, and derives a spectral norm bound that can be used to ensure the positive definiteness of the estimated covariance matrix when sample size is sufficiently large. These results were not available in \citet{zhang2022high}.} 
Under our modeling framework, we further formulate a debiased inferential procedure inspired by the recent literature on debiasing lasso \citep{javanmard2014confidence, zhang2014confidence, cai2022sparse} that can also handle non-Gaussian and heteroskedastic errors. We show that the $q+1$ coefficient matrices associated with $q$ covariates (plus intercept) can be debiased separately, a result that significantly reduces the computational cost.

Although motivated by a biological application, our method provides a general framework for modeling covariance matrices with covariates and is broadly applicable to other scientific fields that involve covariance estimation.

The rest of the paper is organized as follows. Section \ref{sec:meth} introduces the covariance regression model and Section \ref{sec:est} discusses its estimation with sparsity. Section \ref{sec:theory} investigates theoretically the convergence rate of the proposed estimator and also proposes a debiased inferential procedure. Section \ref{sec:simul} carries out comprehensive simulation studies and Section \ref{sec:real} conducts a co-expression QTL analysis using a brain cancer genomics data set. A short discussion section concludes the paper. 



\section{Covariance Regression Models}
\label{sec:meth}
We start with some notation. Write $[d]=\{1,2,\ldots,d\}$. Given a vector $\bx=(x_1, \ldots, x_d)^\top$, we use $\Vert\bx\Vert_1$, $\Vert\bx\Vert_2$ and $\Vert\bx\Vert_{\infty}$ to denote the vector $\ell_1$, $\ell_2$ and $\ell_{\infty}$ norms, respectively.
For a matrix $\bX\in\mathbb{R}^{d_1\times d_2}$, we let $\Vert\bX\Vert_1 = \sum_{ij}|X_{ij}|$, $\Vert\bX\Vert_F=(\sum_{ij}X_{ij}^2)^{1/2}$, $\Vert\bX\Vert_2=\text{sup}_{\bv \neq \bzero} \|\bX \bv\|_2/\|\bv\|_2$ and $\Vert\bX\Vert_{\infty}=\max_{ij}|X_{ij}|$ denote the matrix element-wise $\ell_1$ norm, the Frobenius norm, the spectral norm and the element-wise max norm, respectively. Let $\text{vech}(\bX)=(X_{11}, X_{12},\ldots,X_{1,d_1},\ldots,X_{d_1d_1})$ represents the vectorization of the upper triangular part of $\bX$ and $\text{vec}(\bX)$ represents the concatenation of columns in $\bX$. 
We use $\lambda_{\min}(\cdot)$ and $\lambda_{\max}(\cdot)$ to denote the smallest and largest eigenvalues of a matrix, respectively.

Given a vector of $p$ response variables denoted as $\by=(y_1,\ldots, y_p)^\top$, and a vector of $q$ 
covariates denoted as $\bx=(x_1,\ldots, x_q)^\top$ satisfying $x_l \in [u_l,v_l]$ for $l\in[q]$, we assume that $\mathbb{E}(\by|\bx)=\bbeta_0+\bm\Gamma\bx$, where $\bbeta_0\in\mathbb{R}^{p}$, $\bm\Gamma\in\mathbb{R}^{p\times q}$, and
\be \label{eq:model1}
\text{Cov}(\by|\bx)=\bSigma(\bx) = \bB_0 + \sum_{l=1}^{q} x_l \bB_l,
\ee
where $\bB_0$ is a symmetric and positive definite (PD) matrix of dimension $p \times p$ and $\bB_1,\ldots, \bB_q$ are symmetric matrices of dimension $p \times p$. Here, $\bB_0$ specifies the covariance at the population level and $\bB_l$ represents the effect of covariate $x_l$ on the covariance matrix.
Let $\bB_l=\bQ_l \bLambda_l \bQ_l^\top$ be the eigendecomposition of $\bB_l$ and define $p \times p$ diagonal matrices $\bLambda_l^+$ and $\bLambda_l^-$ such that $\Lambda_{l,jj}^+=\max(0,\Lambda_{l,jj})$ and $\Lambda_{l,jj}^-=\min(0,\Lambda_{l,jj})$, leading to $\bLambda_l=\bLambda_l^+ + \bLambda_l^-$.
\rkcr{We assume that
\be \label{eq:pd_condition}
\lambda_{\text{min}}\bigg\{\bB_0+\sum_{l=1}^q (v_l \bB_l^- + u_l \bB_l^+)\bigg\}>0,
\ee
where $\bB_l^-=\bQ_l \bLambda_l^- \bQ_l^\top$ and $\bB_l^+=\bQ_l \bLambda_l^+ \bQ_l^\top$. 
This is a sufficient condition for a PD $\bSigma(\bx)$ as
$$
\lambda_{\text{min}}\{\bSigma(\bx)\} 
= \lambda_{\text{min}}\bigg\{\bB_0 + \sum_{l=1}^{q} x_l (\bB_l^- + \bB_l^+) \bigg\}
\ge \lambda_{\text{min}}\bigg\{\bB_0 + \sum_{l=1}^{q} (v_l \bB_l^- + u_l \bB_l^+)\bigg\}
$$
where the last inequality holds by the condition $x_l \in [u_l,v_l]$.} 
\rkcr{When $p=1$, we have $\bB_l=b_l$ and condition \eqref{eq:pd_condition} simplifies to $b_0+\sum_{l=1}^{q} \{v_l \min (b_l, 0) + u_l \max (b_l, 0)\} >0$, 
ensuring that the variance remains positive regardless of the values of $x_l$'s. 
For example, in our motivating data example, subjects with a specific genetic variant mutation may have lower variance in gene expression compared to others. In this case, condition \eqref{eq:pd_condition} ensures that these subjects still have a positive variance, as $b_0$ remains dominant.}
We note that requiring the covariates to be bounded in $[u_l,v_l]$ is not restrictive. In our data example, the covariates are genetic variants, which are often coded as $\{0,1\}$. 
To expose key ideas, we assume $\bbeta_0$ and $\bm\Gamma$ are known in the ensuing development, and focus on the estimation of $\bB_0,\bB_1,\ldots,\bB_q$. Extensions with estimated $\bbeta_0$ and $\bm\Gamma$ are straightforward, but with more involved notation.

With $n$ independent observations denoted as $\{(\by_i,\bx_i), i\in[n]\}\in\mathbb{R}^p\times\mathbb{R}^q$, we aim to estimate $\bB_0,\bB_1,\ldots,\bB_q$ via \eqref{eq:model1}. This is a challenging task, as even in the simple Gaussian case, the log likelihood function is
\begin{equation*}
\sum_{i=1}^n\log\left\vert\bB_0+\sum_{l=1}^qx_{il}\bB_l\right\vert-\sum_{i=1}^n\text{tr}\left\{\left(\bB_0+\sum_{l=1}^qx_{il}\bB_l\right)^{-1}\bz_i\bz_i^\top\right\},
\end{equation*}
where $\text{tr}(\cdot)$ denotes the trace of a matrix, $\bz_i=\by_i-\mathbb{E}(\by_i)$, $x_{il}$ is the $l$th element of $\bx_i$ and $z_{ij}$ is the $j$th element of $\bz_i$. 
Due to the sums involved in the matrix trace and inverse calculations, this loglikelihood is not convex or biconvex with respect to $\bB_0,\bB_1,\ldots,\bB_q$, and cannot be directly optimized using iterative algorithmic solutions such as the EM and coordinate descent algorithms. To overcome this challenge, we consider a moment-based approach that is highly efficient to implement and need not to make distributional assumptions on $\by_i$'s. 

First, note that \eqref{eq:model1} implies $\mathbb{E}(z_{ij} z_{ik})=\sum_{l=0}^qx_{il} B_{l, jk}$ with $x_{i0}=1$, that is 
\be\label{eq:model10}
z_{ij} z_{ik} = B_{0, jk} + x_{i1} B_{1, jk} + \ldots + x_{iq} B_{q, jk}+\epsilon_{ijk},
\ee
where $\mathbb{E}(\epsilon_{ijk})=0$ and $B_{l, jk}$ denotes the $(j,k)$th entry of $\bB_l$.  
This observation in \eqref{eq:model10} facilitates the estimation of $\bB_0,\bB_1,\ldots,\bB_q$ via the following least squares estimation
\be \label{eq:lse_ftn}
\sum_{j \leq k} \sum_{i=1}^n (z_{ij} z_{ik} - \sum_{l=0}^q x_{il} B_{l,jk} )^2.
\ee 

When both $p$ and $q$ are large, to ensure the estimability and facilitate the interpretability, 
we impose $\{\bB_0,\bB_1,\ldots,\bB_q\}$ to be sparse. In particular, we assume $\{\bB_1,\ldots,\bB_q\}$ is \textit{group sparse}, corresponding to sparse effective covariates, that is, only a subset of the covariates may impact edges (termed effective covariates).
We further assume each $\bB_l, l \in \{0,1,\ldots,q\}$ is \textit{element-wise sparse}. That is, effective covariates may influence only a subset of the edges. These simultaneous sparsity assumptions are  well supported by genetic studies \citep{gardner2003inferring, vierstra2020global}, and improve model interpretability when compared to using the group sparsity or element-wise sparsity alone. To encourage simultaneous sparsity, we consider the following penalty
\be \label{eq:pen}
\mathcal{P}_{\lambda, \lambda_g}(\bB_0,\bB_1,\ldots,\bB_q)= \lambda \bigg(\sum_{l=1}^q \sum_{j\le k} |B_{l, jk}| + \sum_{j<k} |B_{0, jk}|\bigg) + \lambda_g \sum_{l=1}^q \|\text{vech}(\bB_l)\|_2,
\ee
where 
$\lambda, \lambda_g$ are tuning parameters. 

The term $\sum_{l=1}^q \sum_{j\le k} |B_{l, jk}| + \sum_{j<k} |B_{0, jk}|$ is a lasso penalty that encourages the effect of effective covariates to be sparse. 
The diagonal elements of $\bB_0$ are excluded from element-wise sparse penalty to ensure the response variables have non-zero variances at the population level. 


The term $\sum_{l=1}^q\|\text{vech}(\bB_l)\|_2$ is a group lasso penalty \citep{Yuan2006} that encourages the effective covariates to be sparse. We exclude $\bB_0$ from the group sparse penalty (but not the element-wise sparse penalty), as it determines the population-level covariance matrix. The group sparsity is achieved by regularizing $\bB_l$ across $p(p+1)/2$ regression tasks from \eqref{eq:model10} simultaneously. Correspondingly, this penalty term facilitates a multi-task learning approach \citep{argyriou2008convex}. The penalty term in \eqref{eq:pen} is similar to the sparse group lasso considered in \citet{Simon2013, li2015multivariate}, though it is not exactly the same as some parameters are included in the element-wise sparsity penalty but not the group sparsity penalty. This adds additional complexity to the estimation procedure and theoretical analysis.

\section{Estimation}
\label{sec:est}
Taking into account the condition \eqref{eq:pd_condition}, we consider minimization of the following 
\begin{align}\label{eq:obj_ftn}
& \frac{1}{2n} \sum_{j \leq k} \sum_{i=1}^n (z_{ij} z_{ik} - \sum_{l=0}^q x_{il} B_{l,jk} )^2 + \mathcal{P}_{\lambda, \lambda_g}(\bB_0,\bB_1,\ldots,\bB_q), \nonumber\\
&\quad \text{s.t.}\,\,\lambda_{\text{min}}\bigg\{\bB_0+\sum_{l=1}^q (v_l \bB_l^- + u_l \bB_l^+)\bigg\}>0 
\end{align}
where $x_{i0}=1$ and $\mathcal{P}_{\lambda, \lambda_g}(\cdot)$ is specified as in \eqref{eq:pen}. 
When there are no covariates, \eqref{eq:obj_ftn} reduces to the standard sparse covariance estimation problem \citep{rothman2009generalized, xue2012positive}, written as
\bse
\frac{1}{2n} \sum_{j \leq k} \sum_{i=1}^n (z_{ij} z_{ik} - B_{0, jk})^2 + \lambda \sum_{j < k} |B_{0, jk}|, \quad \text{s.t.} \quad \lambda_{\min}(\bB_0) > 0
\ese
which, assuming $\wh{\bB}_0$ is positive definite, is minimized at $\wh{B}_{0, jk} = S_{\lambda}(\sum_{i=1}^n z_{ij} z_{ik}/n)$ for $j,k\in[p]$ such that $j < k$ and $S_{\lambda}(a)=\text{sign}(a)\times\max(|a|-\lambda, 0)$ is the soft-thresholding operator at $\lambda$ \citep{rothman2009generalized}.
The optimization problem in \eqref{eq:obj_ftn} is nontrivial, as the constraint set is nonconvex. 
\rkcr{Due to the complex form of the constrained optimization problem in \eqref{eq:obj_ftn}, a direct optimization procedure is intractable. Alternatively, we consider an easy-to-compute two-step estimation procedure, which ensures the resulting estimator meets the constraint in \eqref{eq:obj_ftn}, gives an estimator that equals to the solution to \eqref{eq:obj_ftn} asymptotically almost surely, and does not alter the sparsity patterns of the unconstrained estimator.}
The proposed two-step procedure first solves the following non-constrained optimization 
\begin{equation}\label{eq:obj_ftn2}
J(\bB_0,\bB_1,\ldots,\bB_q) = \frac{1}{2n} \sum_{j \leq k} \sum_{i=1}^n (z_{ij} z_{ik} - \sum_{l=0}^q x_{il} B_{l,jk} )^2 + \mathcal{P}_{\lambda, \lambda_g}(\bB_0,\bB_1,\ldots,\bB_q),
\end{equation}
and then adjusts the unconstrained estimator to satisfy the constraint. 
This sequential estimation procedure greatly simplifies the computation and gives asymptotically consistent estimators. Similar sequential procedures have been commonly employed in statistical learning and optimization problems; see, e.g., \citet{li2010generalized,zhang2020mixed}.

\begin{algorithm}[!t]
\caption{Sparse covariance regression with covariates}\label{alg1}
\begin{algorithmic}
\STATE \textbf{Input:} Tuning parameters $\lambda$, $\lambda_g$, convergence tolerance $\xi$, and $\wt{\bB}_l$ as the initial estimator of $\bB_l$ in (\ref{eq:model1}).
\vspace{0.05in}
\REPEAT
\STATE \hspace{0.25in} \textbf{Step 1:} Set $J^{(old)} = J(\wt{\bB}_0,\wt{\bB}_1,\ldots,\wt{\bB}_q)$ 
\STATE \hspace{0.25in} \textbf{Step 2:} For $l=0$, 
compute $\tilde{r}_{ijk}=z_{ij}z_{ik}-\sum_{m \neq l} x_{im} \wt{B}_{m,jk}$ for $i \in [n]$, $j,k \in [p]$ and 
update $\wt{\bB}_0$ for $j,k \in [p]$ by
                \begin{align*}
                \wt{B}_{0,jk} &= \frac{1}{n} \sum_{i=1}^n \tilde{r}_{ijk}  \quad \quad \quad \;\; \text{if  } j=k, \text{ and}\\
			    \wt{B}_{0,jk} &= S_{\lambda}\bigg(\frac{1}{n} \sum_{i=1}^n \tilde{r}_{ijk} \bigg) \quad \text{if  } j \neq k,
			    \end{align*}
where $S_{\lambda}(a)=\text{sign}(a)\times\max(|a|-\lambda, 0)$ is the soft-thresholding operator at $\lambda$.
\vspace{0.05in}
\STATE \hspace{0.25in} \textbf{Step 3:} For $l \in [q]$, compute $\tilde{r}_{ijk}=z_{ij}z_{ik}-\sum_{m \neq l} x_{im} \wt{B}_{m,jk}$ for $i \in [n]$, $j,k \in [p]$ and
check the condition below
                \bse
			    \bigg\|  S_{\lambda}\bigg(\frac{1}{n} \sum_{i=1}^n x_{il} \tilde{\br}_{i,(j \leq k)} \bigg)  \bigg\|_2 < \lambda_g
			    \ese
where $\tilde{\br}_{i,(j \leq k)}$ is the vector of $\tilde{r}_{ijk}$'s for all $j,k$ such that $j \leq k$.
If the condition above is satisfied, set $\wt{\bB}_l=\bzero$.
If not, update $\wt{\bB}_l$ for $j,k \in [p]$ by
			         \begin{align*}
           \wt{B}_{l,jk} = \bigg( \frac{1}{n} \sum_{i=1}^n x_{il}^2 + \frac{\lambda_g}{\| \text{vech}(\wt{\bB}_l) \|_2} \bigg)^{-1} S_{\lambda}\bigg(\frac{1}{n} \sum_{i=1}^n x_{il} \tilde{r}_{ijk} \bigg).
			         \end{align*}
\UNTIL{the algorithm converges: $J^{(old)} - J(\wt{\bB}_0,\wt{\bB}_1,\ldots,\wt{\bB}_q) < \xi$}
\vspace{0.05in}
\STATE \textbf{Step 4:} Compute $\wh\bB_0, \wh{\bB}_1,\ldots, \wh{\bB}_q$ as in \eqref{eq:pd_approx}.
\end{algorithmic}
\end{algorithm}

For the non-constrained optimization \eqref{eq:obj_ftn2}, we adopt the blockwise coordinate descent algorithm as described in Steps 1-3 of Algorithm \ref{alg1}. 
For $l=0$, the solution to the diagonal elements of $\bB_l$ is obtained by the least squares estimator, as the diagonal elements of $\bB_0$ are not penalized, and the solution to the off-diagonal elements of $\bB_l$ is obtained by the lasso estimator, as the off-diagonal elements of $\bB_0$ are not penalized by the group lasso penalty. 
Note that, if the covariates are centered, the solution to $\bB_0$ is equal to the soft thresholding estimator with $\lambda$ as the threshold.
For $l \in [q]$, the solution to $\bB_l$ is obtained by the sparse group lasso estimator. 
Steps 1-3 in Algorithm \ref{alg1} solve the optimization \eqref{eq:obj_ftn2} \rkcr{and convergence of the algorithm is guaranteed by the convergence property of coordinate descent for convex problems with separable penalties \citep{tseng2001convergence}.}
In Step 4 and given the estimators $\widetilde{\bB}_0, \widetilde{\bB}_1,\ldots, \widetilde{\bB}_q$ from Steps 1-3, we set 
\begin{equation} \label{eq:pd_approx}
\wh\bB_0=(1+\delta)^{-1}\widetilde{\bB}_0 + \delta/(1+\delta) \bI_p, \quad\wh\bB_l=(1+\delta)^{-1} \widetilde{\bB}_l, \,\,l \in [q],
\end{equation}
where $\delta = \text{max}[0, -\lambda_{\text{min}}\{\wt{\bB}_0+\sum_{l=1}^q (v_l \wt{\bB}_l^- + u_l \wt{\bB}_l^+) \} ]$.
\rkcr{These final estimates $\wh\bB_0,\wh\bB_1,\ldots, \wh\bB_q$ are easy to compute and uniquely defined by a convex combination of $\wt{\bB}_0+\sum_{l=1}^q (v_l \wt{\bB}_l^- + u_l \wt{\bB}_l^+)$ and $\bI_p$. With these $\wh\bB_0,\wh\bB_1,\ldots, \wh\bB_q$, the covariance matrix $\bSigma(\bx)$ is guaranteed to be positive semi-definite for all possible values of $\bx$ and gives a Ledoit-Wolf type shrinkage estimator \citep{ledoit2004well},
\begin{equation*} 
\wh\bSigma(\bx)=\frac{1}{1+\delta} \bigg(\widetilde{\bB}_0 + \sum_{l=1}^{q} x_l \widetilde{\bB}_l\bigg) + \frac{\delta}{1+\delta} \bI_p.
\end{equation*}
Furthermore, 
$\wh\bB_0,\wh\bB_1,\ldots, \wh\bB_q$ also preserve the sparsity pattern of $\wt{\bB}_0, \wt{\bB}_1,\ldots, \wt{\bB}_q$.}

\rkcr{The final estimates $\wh\bB_0,\wh\bB_1,\ldots, \wh\bB_q$ may not be the exact solution to the constrained optimization \eqref{eq:obj_ftn}. 
However, as $n$ increases, it follows from Theorem \ref{thm2} and Theorem \ref{thm4} that $\widetilde\bB_0, \widetilde\bB_1,\ldots, \widetilde\bB_q$ estimated from Steps 1-3 are consistent and satisfy the PD constraint \eqref{eq:pd_condition} with high probability. That is, as $n$ increases, $\delta$ in \eqref{eq:pd_approx} converges to zero. See more discussion after Theorem \ref{thm4}.}


\rkcr{Two parameters $\lambda$ and $\lambda_g$ in \eqref{eq:obj_ftn2} require tuning. In our procedure, they are jointly selected on a grid of values for $\lambda$ and $\lambda_g$ via $L$-fold cross validation. We let $L=5$ with $\lambda=\alpha \lambda^\ast$ and $\lambda_g=(1-\alpha) \lambda^\ast$ where $\alpha \in \{0.25,0.5,0.75\}$ and $\lambda^\ast \in \{0.01,0.02,\ldots,0.99,1.00\}$ for our simulation studies and real data analysis. More discussion on the computational aspects of Algorithm \ref{alg1} can be found in Section S10 of the Supplementary Materials.}

\section{Theoretical Properties}
\label{sec:theory}


In this section, we first investigate the convergence rate of the estimator from the sparse covariance regression in \eqref{eq:obj_ftn2}, allowing both the response variables and covariates to be high-dimensional.
Next, building on our modeling framework, we develop an inferential procedure using debiasing methodologies.

\subsection{Convergence rate}
Our theoretical analysis on convergence rate encounters new challenges compared to existing work in the literature. 
\rkcr{The penalty term (\ref{eq:pen}) is more involved than a typical sparse group lasso penalty, as $\bB_0$ is excluded from the group sparsity penalty and the diagonal elements of $\bB_0$ are excluded from both the group sparsity penalty and the element-wise sparsity penalties. Particularly, deriving a tight bound for the estimation error of all parameters in our model is challenging because there are $p$ diagonal elements in $\bB_0$, which cannot be assumed to be sparse.} 
\rkcr{In Theorem \ref{thm2}, we show the $\ell_1$-norm and $\ell_2$-norm consistency of all parameters except for the diagonal elements in $\bB_0$. Then, we discuss the spectral norm consistency for our estimator of $\bB_0$ in Theorem \ref{thm4}, which implies that there exists a sufficiently large  $n\ge n_0$ such that the smallest eigenvalue of our estimator of $\bB_0$ will be positive. These results will ultimately ensure that the constrained estimator from \eqref{eq:obj_ftn} reduces to the unconstrained estimator from \eqref{eq:obj_ftn2} when $n$ is sufficiently large (see remark after Theorem \ref{thm4}). Thus, in our theoretical analysis, we will focus on the minimizer of \eqref{eq:obj_ftn2}.}



Let $\bB_0^\ast, \bB_1^\ast,\ldots,\bB_q^\ast$ be the true coefficient matrices in \eqref{eq:model1}. For a matrix $\bB \in \mathbb{R}^{p \times p}$, let $\mathcal{S}(\bB) = \{(j,k): B_{jk} \neq 0, j \leq k\}$ be the index set of non-zero elements in the upper triangular part of $\bB$. Let $|\cdot|$ denote the cardinality of a set. 
Define $s_0=|\mathcal{S}(\bB_0^\ast)|-p$ and $s=s_0 + \sum_{l=1}^q|\mathcal{S}(\bB_0^\ast)|$. That is, $s$ represents the summation of the number of non-zero off-diagonal elements in the upper triangle of $\bB_0^\ast$ and the number of all non-zero elements in the upper triangle of $\bB_1^\ast,\ldots,\bB_q^\ast$. 
Also, define $r=|\{\bB_l^\ast: \bB_l^\ast \neq \bzero, l \in [q]\}|$ as the number of non-zero matrices in $\{\bB_1^\ast,\ldots,\bB_q^\ast\}$.
We first state regularity conditions. 


\begin{Ass} \label{ass:covariate_dist}
Suppose $\wt{\bx}_i$'s are independent and identically distributed (i.i.d) and bounded random vectors with $\mathbb{E}(\wt{\bx}_i)=\bzero$ and a covariance matrix satisfying $\lambda_{\min}\{\mathbb{E}(\wt{\bx}_i \wt{\bx}_i^\top)\}\geq 1/\phi_0$ for some constant $\phi_0 >0$.
Given $n$ observations of $\wt{\bx}_i$, let $\bx_i$ be the centered $\wt{\bx}_i$. Without loss of generality, we assume $|x_{il}|<1$ for all $i$ and $l$. 
\end{Ass}

\begin{Ass} \label{ass:mod}
Model \eqref{eq:model1} holds with $\lambda_{\text{min}}\bigg\{\bB_0^\ast+\sum_{l=1}^q ({\bB_l^\ast}^- - {\bB_l^\ast}^+)\bigg\}>0$
\end{Ass}

\begin{Ass} \label{ass:response_dist}
Suppose $\epsilon_{ijk}$'s are zero-mean sub-exponential random variables, and $\epsilon_{ijk}$ and $\epsilon_{i'j'k'}$ are independent for $i\neq i'$
\end{Ass}



\rkcr{
Assumption \ref{ass:covariate_dist} assumes the centered covariates $\bx_i$ are bounded in $[-1,1]$ and eigenvalues of their covariance matrix are bounded below by a positive constant, that is, $\lambda_{\min}\{\mathbb{E}(\bx_i \bx_i^\top)\}\geq 1/\phi'_0$ for some constant $\phi'_0 >0$.. This is implied by $\lambda_{\min}\{\mathbb{E}(\wt{\bx}_i \wt{\bx}_i^\top)\}\geq 1/\phi_0$, as $\mathbb{E}(\bx_i \bx_i^\top) = (1-n^{-1})\mathbb{E}(\wt{\bx}_i \wt{\bx}_i^\top)$, shown in the proof of Lemma 1.
This condition is not restrictive, 
as one can always rescale the covariates if they are not bounded in $[-1,1]$. Such a transformation of $x_l$ does not affect the interpretability of $\bB_l$; see discussions in Section \ref{sec:conc}.} 
\rkcr{Assumption \ref{ass:mod} describes the parameter constraint \eqref{eq:pd_condition} in our framework with $u_l=-1, v_l=1$ and is reasonable as discussed in Section \ref{sec:meth}.}
\rkcr{Assumption \ref{ass:response_dist} is a condition on the distribution of the response variables. A sufficient condition for this assumption to hold is that the response variables are sub-Gaussian, as the product of two sub-Gaussian random variables is sub-exponential \citep{vershynin2018high}.}

\rkcr{\begin{Th}\label{thm2}
Suppose Assumptions \ref{ass:covariate_dist}, \ref{ass:mod}, and \ref{ass:response_dist} 
hold and assume $s \leq C_1 \sqrt{n/\log \{p(p+1)(q+1)\}}$ for some constant $C_1>0$. 
Denote $\kappa=2^{-1}\min\{(1-n^{-1})\phi_0^{-1}, 1\}$.
Let $\|\epsilon_{ijk}\|_{\psi_1}=\sup_{d \geq 1} d^{-1} (\mathbb{E}|\epsilon_{ijk}|^d)^{1/d}$ be the sub-exponential norm of $\epsilon_{ijk}$ and 
$K = \max_{ijk} \| \epsilon_{ijk}\|_{\psi_1}$. For constants $c>0, C>0$ and $\eta>\max(c^{-1},4)$, let
\bse
\lambda = 2 K \sqrt{ \frac{\eta \log \{p (p+1) (q+1)\}}{n} } \quad \text{and} \quad \lambda_g = C \lambda \sqrt{\frac{s}{r}}.
\ese
If $C_1 = \kappa/(36 \sqrt{\eta})$, the solution $\wh{\bB}_0,\ldots,\wh{\bB}_q$ to the optimization \eqref{eq:obj_ftn2} satisfy
\begin{equation*}
\bigg\{ \sum_{j=k}\sum_{l=1}^q (B_{l, jk}^\ast-\wh{B}_{l,jk})^2 + \sum_{j<k}\sum_{l=0}^q (B_{l, jk}^\ast-\wh{B}_{l,jk})^2 \bigg\}^{1/2} \leq \frac{(3 + 2 C)\sqrt{s}\lambda}{\kappa},
\end{equation*}
and
\begin{equation*}
\sum_{j=k}\sum_{l=1}^q |B_{l, jk}^\ast-\wh{B}_{l,jk}| + \sum_{j<k}\sum_{l=0}^q |B_{l, jk}^\ast-\wh{B}_{l,jk}| \leq \frac{4(2+C)^2 \lambda s}{\kappa}
\end{equation*}
with probability at least $1 - 3\{p(p+1)(q+1)\}^{\max(1-c \eta, 2-\eta/2)}$.
\end{Th}}

Theorem \ref{thm2} shows that the estimation error of our estimator except for the diagonal elements of $\bB_0$ is bounded by a factor of order $\sqrt{s\max(\log p,\log q)/n}$.
Here, we do not assume $\epsilon_{ijk}$ and $\epsilon_{ij'k'}$ are independent, as $z_{ij}z_{ik}$ and $z_{ij'}z_{ik'}$ from gene pairs $(j,k)$ and $(j',k')$ can be correlated. 
Hence, under our setting, the error terms across element-wise regression tasks can be correlated. Comparable convergence rates, up to a logarithm factor, have been derived for sparse group lasso estimators in univariate regressions \citep{cai2022sparse,zhang2022high}. 
From the $\ell_1$-norm estimation error in Theorem \ref{thm2}, the following result for the estimation error of individuals' covariance matrices is also available.

\rkcr{
\begin{Cor} \label{cor1}
Suppose assumptions in Theorem \ref{thm2} hold. Denote $\wh{\bSigma}(\bx) = \wh{\bB}_0 + \sum_{l=1}^{q} x_{l} \wh{\bB}_l$ and let $\bSigma^\ast(\bx) = \bB_0^\ast + \sum_{l=1}^{q} x_{l} \bB_l^\ast$ be the true covariance matrix. 
Then,
\bse
\sum_{j<k}|\wh{\Sigma}_{jk}(\bx)-\Sigma_{jk}^\ast(\bx)| \leq \frac{4(2+C)^2 \lambda s}{\kappa}
\ese
with probability at least $1 - 3\{p(p+1)(q+1)\}^{\max(1-c \eta, 2-\eta/2)}$ where $\wh{\Sigma}_{jk}(\bx)$ and $\Sigma_{jk}^\ast(\bx)$ are the $(j,k)$th element of $\wh{\bSigma}(\bx)$ and $\bSigma^\ast(\bx)$, respectively.
\end{Cor}}

\rkcr{Next, we establish a spectral norm bound on the convergence rate for $\wh{\bB}_0$.} 

\rkcr{\begin{Th}\label{thm4}
Suppose Assumptions in Theorem \ref{thm2} hold, 
and we have 
\bse
\|\wh{\bB}_0-\bB_0^\ast\|_2 = O_P \bigg\{ (s_0+1) \bigg(\frac{\log p(p+1)(q+1)}{n}\bigg)^{\frac{1}{2}} \bigg\}.
\ese
\end{Th}}
\rkcr{Theorem \ref{thm2} and Theorem \ref{thm4} suggests that solutions $\wt{\bB}_0, \wt{\bB}_1,\ldots,\wt{\bB}_q$ to the optimization \eqref{eq:obj_ftn2} satisfy the PD constraint in \eqref{eq:obj_ftn} as $n$ increases.
Specifically, it holds by Weyl's inequality that
\begin{align*}
\lambda_{min}\bigg\{\bB_0^\ast+\sum_{l=1}^q ({\bB_l^{\ast}}^- - {\bB_l^{\ast}}^+)\bigg\} 
&\leq \lambda_{min}\bigg\{\wt{\bB}_0+\sum_{l=1}^q (\wt{\bB}_l^- - \wt{\bB}_l^+)\bigg\}\\
& \quad - \lambda_{min}\bigg\{\wt{\bB}_0+\sum_{l=1}^q (\wt{\bB}_l^- - \wt{\bB}_l^+) -\bB_0^\ast - \sum_{l=1}^q ({\bB_l^{\ast}}^- - {\bB_l^{\ast}}^+)\bigg\}\\
&\leq \lambda_{min}\bigg\{\wt{\bB}_0+\sum_{l=1}^q (\wt{\bB}_l^- - \wt{\bB}_l^+)\bigg\}\\
& \quad - \lambda_{min}(\wt{\bB}_0 -\bB_0^\ast) - \lambda_{min}\bigg\{\sum_{l=1}^q (\wt{\bB}_l^- - \wt{\bB}_l^+) - \sum_{l=1}^q ({\bB_l^{\ast}}^- - {\bB_l^{\ast}}^+)\bigg\}.
\end{align*}
Since $\lambda_{min}\{\sum_{l=1}^q (\wt{\bB}_l^- - \wt{\bB}_l^+) - \sum_{l=1}^q ({\bB_l^{\ast}}^- - {\bB_l^{\ast}}^+)\}$ and $ \lambda_{min}(\wt{\bB}_0 -\bB_0^\ast)$ converge to zero by Theorem \ref{thm2} and Theorem \ref{thm4}, respectively, and $\lambda_{\text{min}}\{\bB_0^\ast+\sum_{l=1}^q ({\bB_l^{\ast}}^- - {\bB_l^{\ast}}^+) \}>0$ by Assumption \ref{ass:mod}, we have, 
\bse
\lambda_{\text{min}}\bigg\{\wt{\bB}_0+\sum_{l=1}^q (\wt{\bB}_l^- - \wt{\bB}_l^+)\bigg\}>0.
\ese
for a sufficiently large sample size $n_0$.
That is, the unconstrained optimizer of \eqref{eq:obj_ftn2} satisfies the PD constraint in \eqref{eq:obj_ftn} for any $n \geq n_0$, and consequently, the constrained and unconstrained optimizers are asymptotically equal almost surely.}

\subsection{Statistical Inference via Debiasing}
\label{sec:inference}
We consider the inference for $\bB_0^\ast,\bB_1^\ast,\ldots,\bB_q^\ast$ under the proposed sparse covariance regression. 
\rkcr{We employ debiasing methodologies for statistical inference instead of performing the inference directly on the estimator, as lasso-type estimators do not admit exact characterization of asymptotic limits and suffer from non-negligible bias, leading to inaccurate results in inference \citep{javanmard2014confidence, zhang2014confidence}.}
Hence, inspired by recent advances on debiasing in high dimensional linear regressions \citep{javanmard2014confidence, zhang2014confidence,cai2022sparse}, we design a debiased lasso estimator which performs debiasing on each vector $\wh{\bB}_{\mydot,jk}$ for $j,k \in [p]$ separately and make inference on the true parameter matrices $\bB_0^\ast,\bB_1^\ast,\ldots,\bB_q^\ast$. Since $\{\wh{\bB}_0,\wh{\bB}_1,\ldots,\wh{\bB}_q\}$ is of dimension $p^2(q+1)$, carrying out the debiasing each vector $\wh{\bB}_{\mydot,jk}$ of dimension $q+1$ significantly reduces the computational cost. The cost for this computational gain is the potential loss of efficiency, compared to debiasing all elements in $\{\wh{\bB}_0,\wh{\bB}_1,\ldots,\wh{\bB}_q\}$ simultaneously. We also note that debiasing all elements in $\{\wh{\bB}_0,\wh{\bB}_1,\ldots,\wh{\bB}_q\}$ simultaneously may not be feasible under our framework as one needs to estimate $\text{Cov}(z_{ij}z_{ik}, z_{ij'}z_{ik'})$, the analytical form of which is difficult to derive without distributional assumptions on $\bz_{i}$.
Besides non-Gaussian errors, our procedure also faces the challenge of heteroskedasticity, as $\text{Var}(\epsilon_{ijk})$ may depend on $\bx_i$. Next, we detail our approach. 

Let $\bX=\{x_{il}\}_{i=1,l=0}^{n,q}$ is the $n \times (q+1)$ design matrix (including the intercept term) and denote $\wh{\bTheta} = \bX^\top \bX /n$. 
For $j,k \in [p]$ such that $j \leq k$, consider the following debiased estimator $\wh{\bB}_{\mydot,jk}^u$:
\be \label{eq:debias}
\wh{\bB}_{\mydot,jk}^u = \wh{\bB}_{\mydot,jk} + \frac{1}{n} \bM \bX^\top ( \bz_{\mydot j} \circ \bz_{\mydot k} - \bX \wh{\bB}_{\mydot,jk})
\ee
where $\bz_{\mydot j} \in \mathbb{R}^n$ is a vector of $z_j$ for all $i \in [n]$, $\circ$ denotes the element-wise product of two equal-length vectors, $\bM=[\m_0,\ldots,\m_q]^\top$, and $\m_l$ for $l\in \{0,1,\ldots,q\}$ is a solution to 
\begin{align} \label{eq:debias_optim}
&\m_l=\arg\min_{\m} \m^\top \wh{\bTheta} \m \nonumber \\
&\text{subject to} \quad  \| \wh{\bTheta}\m - \b1e_l \|_\infty \leq \mu \nonumber, \\
& \quad \quad \quad \quad \quad \;  \| \bX \m \|_{\infty} \leq n^{\beta}, \quad \text{for any fixed} \quad 1/4 < \beta < 1/2,
\end{align}
where $\mu$ is to be specified later and $\b1e_l$ is the $(l+1)$th vector in the canonical basis of $\mathbb{R}^{q+1}$. From the above calculations, $\bM$ is only a function of the design matrix $\bX$, and can be seen as an approximation to the inverse of $\bX^\top \bX /n$. 
\rkcr{The feasibility of the optimization \eqref{eq:debias_optim} is ensured by Assumption \ref{ass:covariate_dist} and results in \citet{javanmard2014confidence}. }

\begin{Th}\label{thm3}
Suppose assumptions in Theorem \ref{thm2} hold and optimizations in \eqref{eq:debias_optim} are feasible. Assume $\mathbb{E}(\epsilon_{ijk}^2)=\sigma_{ijk}^2$ and $\mathbb{E}(|\epsilon_{ijk}|^{2+a})<C_2\sigma_{ijk}^{2+a}$ for some $a>0$ and $C_2>0$. 
\begin{itemize}
\item[(1)] For $l \in \{0,1,\ldots,q\}$, with probability at least $1 - 3\{p(p+1)(q+1)\}^{\max(1-c \eta, 2-\eta/2)}$, $\wh{\bB}_l^u$ can be decomposed by $\bLambda_l, \bW_l \in \mathbb{R}^{p \times p}$ as 
\bse
\sqrt{n}(\wh{\bB}_l^u-\bB_l^\ast)= \bLambda_l + \bW_l,
\ese
where $\Lambda_{l,jk} = n^{1/2} (\m_l^\top \wh{\bTheta} - \b1e_l^\top) (\bB_{\mydot,jk}^\ast - \wh{\bB}_{\mydot,jk})$ such that
\bse
\max_{\substack{j < k \\l \in \{0,1,\ldots,q\}}} |\Lambda_{l,jk}|  \leq \frac{8(2+C)^2 K}{\kappa} \mu s \sqrt{ \eta \log \{p (p+1) (q+1)\} },
\ese
and $W_{l,jk} = n^{-1/2} \sum_{i=1}^n (\bX \m_l)_i \epsilon_{ijk}$ is asymptotically normal with mean zero and variance $n^{-1}\sum_{i=1}^n \{(\bX \m_l)_i\}^2 \sigma_{ijk}^2$. 

\item[(2)] When $\mu=\sqrt{ \log \{p (p+1) (q+1)\}/n }$ and $s\log \{p (p+1) (q+1)\}/\sqrt{n}=o(1)$, 
an asymptotic two-sided $100(1-\alpha)\%$ confidence interval for $B_{l,jk}^\ast$ for $j < k$ is
\bse
\wh{B}_{l,jk}^u \pm \Phi^{-1}(1-\alpha/2) n^{-1} \sqrt{\sum_{i=1}^n \{(\bX \m_l)_i\}^2 \sigma_{ijk}^2}.
\ese
\end{itemize}
\end{Th}

In Theorem \ref{thm3}, the errors are not assumed to be Gaussian and they can be heteroskedastic. Calculating the above confidence interval requires estimating the variance of $W_{l,jk}$. This can be consistently estimated by the empirical variance \citep{buhlmann2015high}
\be \label{eq:var_est}
\frac{1}{n} \sum_{i=1}^n \bigg\{ (\bX \m_l)_i \hat{\epsilon}_{ijk}-\frac{1}{n} \sum_{h=1}^n (\bX \m_l)_h \hat{\epsilon}_{hjk} \bigg\}^2.
\ee
where $\hat{\epsilon}_{ijk}=z_{ij}z_{ik} - \sum_{l=0}^q x_{il} \wh{B}_{l,jk}^u$.

\section{Simulation Studies}
\label{sec:simul}
In this section, we investigate the finite sample performance of our proposed method, referred to as \texttt{SparseCovReg}, and compare it with four alternative methods, including:\\
$\bullet$ \texttt{DenseSample}: standard sample covariance estimator $\bS = \sum_{i=1}^n \bz_i \bz_i^\top /n$,\\
$\bullet$ \texttt{SparseSample}: soft-thresholding sample covariance estimator $S_\lambda(\bS)$ where $S_\lambda(\cdot)$ is \\
\text{     }the element-wise soft-thresholding operator at $\lambda$ \citep{rothman2009generalized},\\
$\bullet$ \texttt{CovReg}: quadratic covariance regression estimator in \citet{hoff2012covariance},\\
$\bullet$ \texttt{DenseCovReg}: the least squares estimator obtained by minimizing \eqref{eq:lse_ftn}.\\
The tuning parameters in \texttt{SparseCovReg} and \texttt{SparseSample} are selected using 5-fold cross validation.


We simulate $n$ samples $\{(\by_i,\bx_i), i\in[n]\}$, where the response $\by_i$ is of dimension $p$ (e.g., genes) and covariate $\bx_i$ is of dimension $q$ (e.g., genetic variants). 
For $\bx_i$'s, we consider two simulation settings. In {\bf Setting 1,} we consider continuous covariates drawn independently from $\text{Uniform}(0,1)$ and in {\bf Setting 2}, we consider discrete covariates drawn independently from $\text{Bernoulli}(0.5)$.
\rkcr{Given $\bx_i$, we simulate $\by_i$ from $N_{p}(\bzero, \bSigma(\bx_i))$, where the covariance matrix follows three different types of structures. In particular, we consider, for $j \leq k$,
\begin{itemize}
\item MA(1): $\Sigma_{jk}(\bx)=
\begin{cases}
    0.5 + 0.5 x_1,       & \quad \text{if } j=k,\\
    0.5 x_1,  & \quad \text{if } |j-k|=1,\\
    0 ,       & \quad \text{if } |j-k|>1,
\end{cases}$
\item Clique: 
\begin{align*}
& \Sigma(\bx)=
\begin{bmatrix}
\bTheta(\bx) & \bzero & \bzero & \bzero & \bzero \\
\bzero & \bTheta(\bx) & \bzero & \bzero & \bzero \\
\bzero & \bzero & \bTheta(\bx) & \bzero & \bzero \\
\bzero & \bzero & \bzero & \bTheta(\bx) & \bzero \\
\bzero & \bzero & \bzero & \bzero & \bTheta(\bx)
\end{bmatrix}, 
\end{align*}
where $\bTheta(\bx)$ is of size $10\times 10$, the diagonal elements are equal to $0.5 + 0.5 x_1$ and the off-diagonal elements are equal to $0.5 x_1$.

\item Hub: $\Sigma_{jk}(\bx)=
\begin{cases}
    0.5 + 0.5 x_1,       & \quad \text{if } j=k,\\
    0.4 x_1,  & \quad \text{if } $\text{mod}(j,5)=1, $k \in \{j+1,\ldots,j+4\}$$\\
    0 ,       & \quad \text{otherwise}.
\end{cases}$
\end{itemize}}
These covariance structures have been commonly considered by others \citep{rothman2009generalized, bien2011sparse, qiu2019threshold, xu2022proximal}. 
We consider $n=200,500$, $p=50$ and $q=30,100$. For each simulation configuration, we generate 100 independent data sets.

\begin{figure}
\centering
\scalebox{0.37}{
\includegraphics{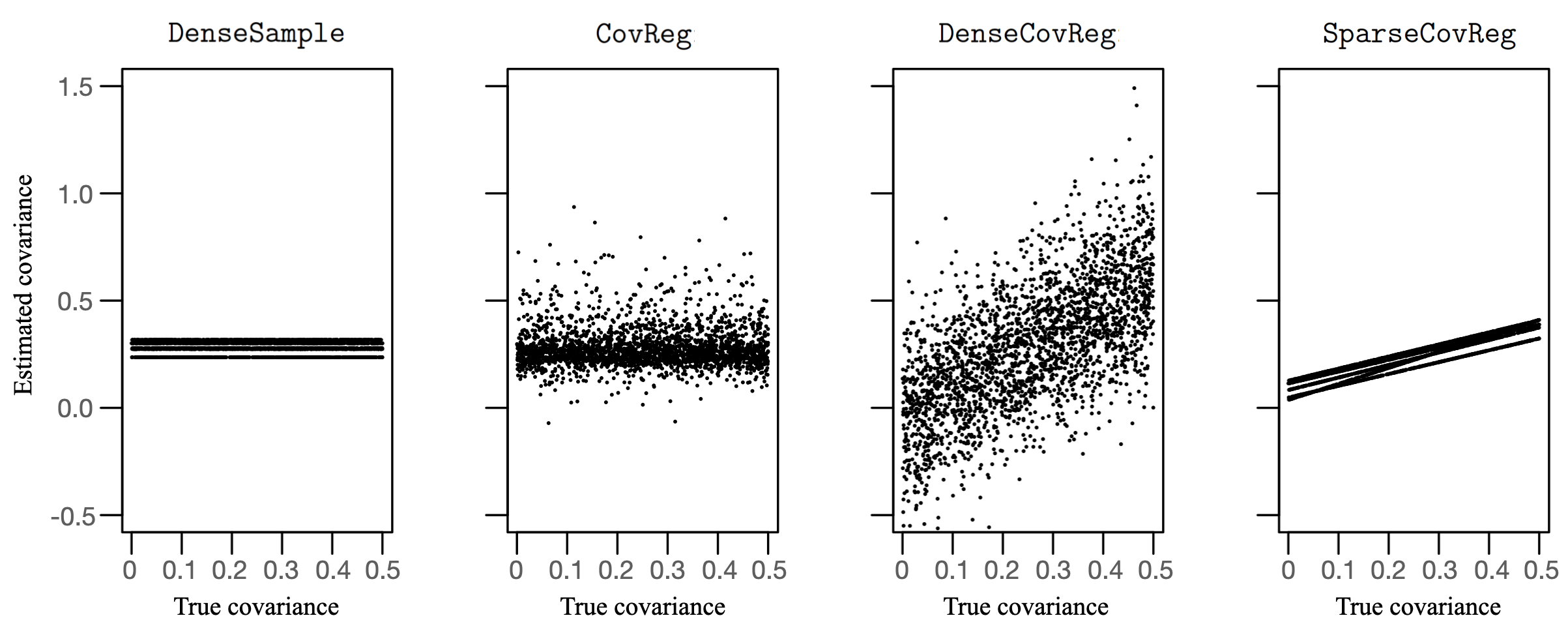}
}
\caption{Comparison of the true non-zero covariance $\Sigma_{i,12}^\ast$ (x-axis) and estimated covariance $\wh{\Sigma}_{i,12}$ (y-axis) for five simulated datasets from the MA(1) model under Setting 1 (continuous covariates) with the number of responses $p=50$, the number of covariates $q=30$ and the sample size $n=500$.}
\label{fig:scatter}
\end{figure}

Let $\bSigma_i^\ast$ denotes the true covariance matrix for the $i$th observation and $\wh{\bSigma}_i$ denotes the estimated $\bSigma_i^\ast$ from a given method. 
For illustration, Figure \ref{fig:scatter} plots $\wh{\Sigma}_{i,12}$, the $(1,2)$th entry of $\wh{\Sigma}_i$, against $\Sigma_{i,12}^\ast$ for $i\in[n]$ from 5 data replicates. 
We did not include the scatter plot from \texttt{SparseSample} as it is very similar to that of \texttt{DenseSample}. As \texttt{DenseSample} does not account for the effect of covariates, $\Sigma_{i,12}^\ast$ is estimated to be constant across all subjects, as shown by five horizontal lines from 5 data replicates. 
The \texttt{CovReg} method by \citet{hoff2012covariance} cannot estimate the covariance well as the true covariance is not a quadratic function of the covariates. 
The \texttt{DenseCovReg} gives a reasonable agreement between the estimated and true covariances (slopes are all roughly 1), though the variability of the estimates is very high.
On the other hand, the proposed \texttt{SparseCovReg} estimates the covariance well (slopes are all roughly 1) and enjoys a much reduced variability. 

\begin{table}[!t]
\centering
\scalebox{0.8}{
{\renewcommand{\arraystretch}{0.75}
\color{black}
\begin{tabular}{ccccccccc}
  \hline \hline
   &  &  & \multicolumn{2}{c}{\uline{MA(1)}} & \multicolumn{2}{c}{\uline{Clique}} & \multicolumn{2}{c}{\uline{Hub}} \\[1.0ex]
   $n$ & $q$ & method & Setting 1 & Setting 2 & Setting 1 & Setting 2 & Setting 1 & Setting 2 \\[0.5ex] \hline
   200 & 30 & \texttt{DenseSample} & 3.22 (0.08) & 4.15 (0.07) & 4.11 (0.12) & 6.26 (0.09) & 3.08 (0.08) & 3.78 (0.07) \\ [1.0ex]
     &  & \texttt{SparseSample} & 2.01 (0.06) & 3.23 (0.06) & 3.51 (0.12) & 5.84 (0.10) & 1.77 (0.05) & 2.73 (0.05) \\ [1.0ex]
     &  & \texttt{CovReg} & 7.81 (0.34) & 8.65 (0.39) & 8.26 (0.42) & 9.73 (0.41) & 7.93 (0.34) & 9.01 (0.49) \\ [1.0ex]
     &  & \texttt{DenseCovReg} & 15.09 (0.34) & 15.64 (0.46) & 15.16 (0.38) & 15.76 (0.54) & 15.07 (0.30) & 15.61 (0.43) \\ [1.0ex]
     &  & \texttt{SparseCovReg} & \textbf{1.87} (0.06) & \textbf{2.14} (0.09) & \textbf{3.30} (0.20) & \textbf{3.53} (0.28) & \textbf{1.69} (0.06) & \textbf{1.99} (0.08) \\ [1.0ex] \hline
     & 100 & \texttt{DenseSample} & 3.21 (0.08) & 4.13 (0.08) & 4.10 (0.13) & 6.26 (0.10) & 3.06 (0.07) & 3.76 (0.07) \\ [1.0ex]
     &  & \texttt{SparseSample} & 2.01 (0.06) & 3.22 (0.06) & 3.50 (0.12) & 5.84 (0.09) & 1.76 (0.05) & 2.72 (0.05)  \\ [1.0ex]
     &  & \texttt{CovReg} & 14.45 (1.75) & 15.14 (1.61) & 14.73 (1.69) & 16.05 (1.70) & 14.45 (1.82) & 17.34 (2.49)  \\ [1.0ex]
     &  & \texttt{DenseCovReg} & 26.92 (0.60) & 27.64 (0.79) & 27.06 (0.78) & 27.87 (1.06) & 26.91 (0.57) & 27.63 (0.75)  \\ [1.0ex]
     &  & \texttt{SparseCovReg} & \textbf{1.89} (0.07) & \textbf{2.19} (0.12) & \textbf{3.37} (0.22) & \textbf{3.63} (0.35) & \textbf{1.70} (0.07) & \textbf{2.01} (0.10) \\ [1.0ex] \hline
    500 & 30 & \texttt{DenseSample} & 2.40 (0.04) & 3.53 (0.03) & 3.44 (0.08) & 5.87 (0.04) & 2.22 (0.04) & 3.08 (0.03) \\ [1.0ex]
     &  & \texttt{SparseSample} & 1.75 (0.04) & 3.12 (0.03) & 3.13 (0.07) & 5.70 (0.04) & 1.50 (0.03) & 2.60 (0.02) \\ [1.0ex]
     &  & \texttt{CovReg} & 4.39 (0.11) & 5.28 (0.17) & 5.13 (0.19) & 7.04 (0.15) & 4.39 (0.17) & 5.17 (0.22) \\ [1.0ex]
     &  & \texttt{DenseCovReg} & 9.55 (0.14) & 9.93 (0.18) & 9.64 (0.18) & 10.07 (0.23) & 9.54 (0.13) & 9.92 (0.18) \\ [1.0ex]
     &  & \texttt{SparseCovReg} & \textbf{1.29} (0.06) & \textbf{1.39} (0.07) & \textbf{2.19} (0.17) & \textbf{2.23} (0.17) & \textbf{1.19} (0.05) & \textbf{1.30} (0.06) \\ [1.0ex] \hline
     & 100 & \texttt{DenseSample} & 2.39 (0.04) & 3.52 (0.03) & 3.44 (0.07) & 5.86 (0.04) & 2.21 (0.04) & 3.08 (0.03) \\ [1.0ex]
     &  & \texttt{SparseSample} & 1.75 (0.04) & 3.11 (0.03) & 3.12 (0.07) & 5.69 (0.04) & 1.50 (0.03) & 2.60 (0.02) \\ [1.0ex]
     &  & \texttt{CovReg} & 11.12 (0.34) & 12.22 (0.44) & 11.34 (0.39) & 12.76 (0.43) & 11.49 (0.33) & 13.42 (0.47) \\ [1.0ex]
     &  & \texttt{DenseCovReg} & 17.26 (0.28) & 17.86 (0.36) & 17.42 (0.35) & 18.07 (0.46) & 17.26 (0.25) & 17.87 (0.33) \\ [1.0ex]
     &  & \texttt{SparseCovReg} & \textbf{1.32} (0.06) & \textbf{1.43} (0.07) & \textbf{2.26} (0.20) & \textbf{2.35} (0.21) & \textbf{1.21} (0.05) & \textbf{1.33} (0.06) \\ [1.0ex]
  \hline \hline
\end{tabular}}}
\caption{\rkcr{Average error for individual covariance matrix measured by $n^{-1} \sum_{i=1}^n \|\wh{\bSigma}_i-\bSigma_i^\ast\|_F$ over 100 simulations with standard error shown in parentheses. The lowest error in each setting has been bolded.}} 
\label{tb:frob}
\end{table}

Next, we compare the average error for all individuals' covariance matrices measured by $n^{-1} \sum_{i=1}^n \|\wh{\bSigma}_i-\bSigma_i^\ast\|_F$. 
Table \ref{tb:frob} reports the average errors with standard errors in the parentheses. 
The proposed \texttt{SparseCovReg} outperforms the alternative methods for all $n$ and $q$. It is seen that the error of \texttt{SparseCovReg} decreases with $n$ and slightly increases with $q$, confirming the results of Theorem \ref{thm2} and Corollary \ref{cor1}. 

\rkcr{Additionally, in Table \ref{tb:tprfpr}, we summarize the performance of \texttt{SparseCovReg} by reporting the root sum of squared error (RSSE) of $\wh{\bB}_0,\wh{\bB}_1,\ldots,\wh{\bB}_q$ 
\begin{equation*} 
\text{RSSE} = \bigg\{\sum_{j \leq k}\sum_{l=0}^q (B_{l, jk}^\ast-\wh{B}_{l,jk})^2 \bigg\}^{1/2},
\end{equation*}
and the true positive rate (TPR) and the false positive rate (FPR)
\begin{align*}
\text{TPR} & = \frac{\#\{(l,j,k): \wh{B}_{l,jk} \neq 0,\; B_{l,jk}^\ast \neq 0 \}}{\#\{(l,j,k): B_{l,jk}^\ast \neq 0 \}}, \\
\text{FPR} & = \frac{\#\{(l,j,k): \wh{B}_{l,jk} \neq 0,\; B_{l,jk}^\ast = 0 \}}{\#\{(l,j,k): B_{l,jk}^\ast = 0\}}.
\end{align*}
Similar to Table \ref{tb:frob}, RSSE decreases with $n$ and slightly increases with $q$, confirming the results of Theorem \ref{thm2}.}
Note that the selection accuracy cannot be fairly evaluated from other methods, as \texttt{DenseSample}, \texttt{CovReg} and \texttt{DenseCovReg} are all dense estimators, and \texttt{SparseSample} does not estimate $B_{l,jk}$ for $l\in[q]$.

\begin{table}[!t]
\centering
\scalebox{0.9}{
{\renewcommand{\arraystretch}{0.75}
\color{black}
\begin{tabular}{ccccccccc}
  \hline \hline \\ [-1.5ex] 
  &  & Performance & \multicolumn{2}{c}{\uline{MA(1)}} & \multicolumn{2}{c}{\uline{Clique}} & \multicolumn{2}{c}{\uline{Hub}} \\[1.0ex]
   $n$ & $q$ & measure & Setting 1 & Setting 2 & Setting 1 & Setting 2 & Setting 1 & Setting 2 \\[1.0ex] \hline 
   200 & 30 & RSSE & 4.5820 & 3.2573 & 6.9493 & 4.7021 & 4.1801 & 3.1175 \\ [1.0ex]
    &  &  & (0.1651) & (0.1195) & (0.3972) & (0.3283) & (0.1388) & (0.1193) \\ [1.0ex]
    &  & TPR & 0.7662 & 0.9865 & 0.7729 & 0.9937 & 0.6767 & 0.9563 \\ [1.0ex]
     &  & FPR & 0.0076 & 0.0100 & 0.0445 & 0.0629 & 0.0057 & 0.0076 \\ [1.0ex] \hline 
     & 100 & RSSE & 4.6364 & 3.3434 & 7.1308 & 4.8968 & 4.2472 & 3.1815 \\ [1.0ex]
     &  &  & (0.1787) & (0.1660) & (0.4504) & (0.4517) & (0.1783) & (0.1378) \\ [1.0ex]
     &  & TPR & 0.7520 & 0.9846 & 0.7349 & 0.9902 & 0.6567 & 0.9552 \\ [1.0ex]
     &  & FPR & 0.0016 & 0.0019 & 0.0079 & 0.0121 & 0.0013 & 0.0016 \\ [1.0ex] \hline 
    500 & 30 & RSSE & 3.4257 & 2.1246 & 4.9594 & 3.0079 & 3.2422 & 2.0567 \\ [1.0ex]
     &  &  & (0.1987) & (0.1075) & (0.4847) & (0.2356) & (0.1726) & (0.0989) \\ [1.0ex]
     &  & TPR & 0.9742 & 1.0000 & 0.9849 & 1.0000 & 0.9377 & 0.9998 \\ [1.0ex]
     &  & FPR & 0.0107 & 0.0129 & 0.0669 & 0.0757 & 0.0080 & 0.0095 \\ [1.0ex] \hline 
     & 100 & RSSE & 3.5304 & 2.2072 & 5.2053 & 3.1960 & 3.3370 & 2.1180 \\ [1.0ex]
     &  &  & (0.2105) & (0.1156) & (0.4888) & (0.2914) & (0.1690) & (0.1048) \\ [1.0ex]
     &  & TPR & 0.9743 & 1.0000 & 0.9800 & 1.0000 & 0.9306 & 1.0000 \\ [1.0ex]
     &  & FPR & 0.0018 & 0.0021 & 0.0120 & 0.0137 & 0.0015 & 0.0016 \\ [1.0ex]
  \hline \hline
\end{tabular}}}
\caption{\rkcr{Average root sum of squared error (RSSE) with standard error shown in parentheses, true positive rate (TPR) and false positive rate (FPR) of \texttt{SparseCovReg} over 100 simulations.} 
}
\label{tb:tprfpr}
\end{table}

\begin{figure}[!t]
\centering
\begin{overpic}[width=5.9in,angle=0]{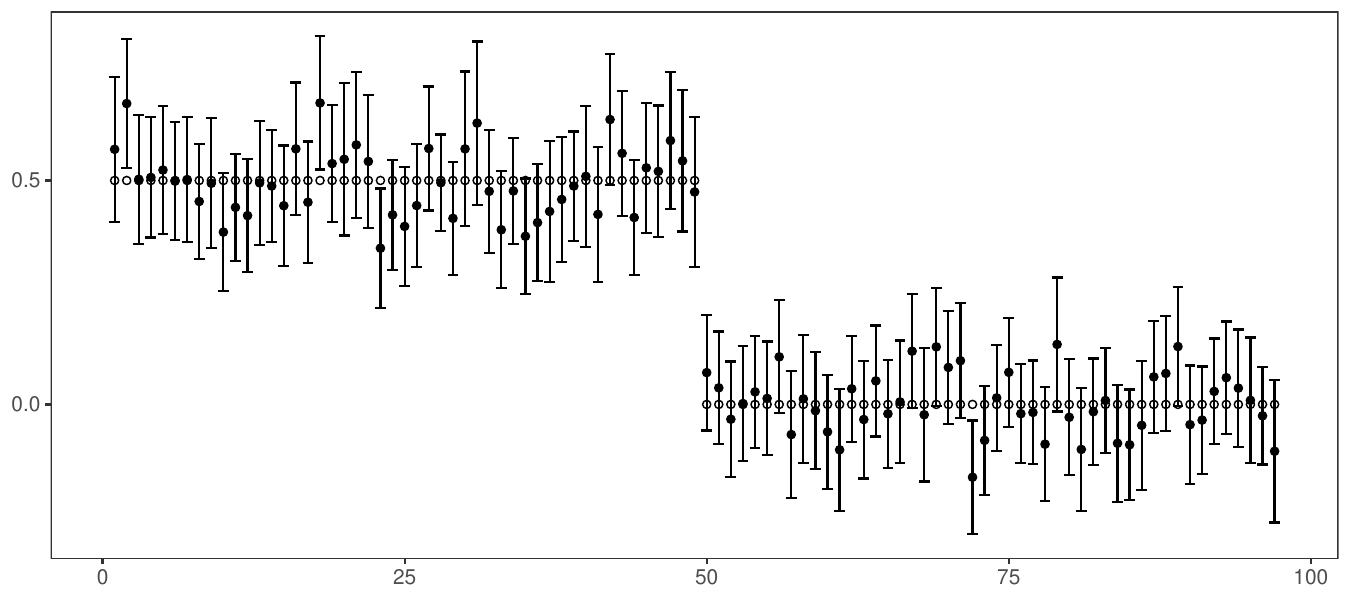}
\put(48,-2){\small Index}
\put(-2.0,18){\rotatebox{90}{\small{Coefficient}}}
\end{overpic}
\vspace*{6mm}
\caption{95\% confidence intervals for elements in $\bB_1^\ast$ from one data replicate from the MA(1) model under Setting 2 (binary covariates) with the number of responses $p=50$, the number of covariates $q=100$ and the sample size $n=500$. True parameter values are shown in $\circ$ and estimated parameter values, after debiasing, are shown in {\small $\bullet$}.}
\label{fig:ciplot}
\end{figure}

 \begin{table}[!t]
\centering
\scalebox{0.9}{
\color{black}
\begin{tabular}{cccccccccc}
  \hline \hline
& &  & & \multicolumn{3}{c}{\uline{Setting 1}} & \multicolumn{3}{c}{\uline{Setting 2}}\\[0.7ex]
 & $n$ & $q$ & Method & $\bB_l^\ast$ & $\mathcal{S}(\bB_l^\ast)$ & $\mathcal{S}^c(\bB_l^\ast)$ & $\bB_l^\ast$ & $\mathcal{S}(\bB_l^\ast)$ & $\mathcal{S}^c(\bB_l^\ast)$ \\[1.0ex] \hline
     MA(1) & 200 & 30 & $\hat\sigma_{ijk}$& 0.938 & 0.927 & 0.938 & 0.927 & 0.910 & 0.927  \\ [1.0ex]
     & &  &$\sigma_{ijk}^\ast$& 0.950 & 0.948 & 0.950 & 0.950 & 0.948 & 0.950\\[1.0ex] \cline{3-10}
     & & 100 &$\hat\sigma_{ijk}$& 0.876 & 0.856 & 0.876 & 0.864 & 0.830 & 0.864  \\ [1.0ex]
     & &  &$\sigma_{ijk}^\ast$& 0.951 & 0.947 & 0.951 & 0.951 & 0.934 & 0.951\\[1.0ex]\cline{2-10}
     & 500 & 30 &$\hat\sigma_{ijk}$ & 0.943 & 0.936 & 0.943 & 0.942 & 0.939 & 0.942  \\ [1.0ex]
     & &  &$\sigma_{ijk}^\ast$& 0.951 & 0.947 & 0.951 & 0.951 & 0.953 & 0.951\\[1.0ex]\cline{3-10}
     & & 100 &$\hat\sigma_{ijk}$& 0.925 & 0.912 & 0.925 & 0.921 & 0.909 & 0.921  \\ [1.0ex]
     & &  &$\sigma_{ijk}^\ast$& 0.951 & 0.947 & 0.951 & 0.951 & 0.951 & 0.951\\[1.0ex]\hline
     Clique & 200 & 30 & $\hat\sigma_{ijk}$& 0.938 & 0.927 & 0.938 & 0.927 & 0.912 & 0.927  \\ [1.0ex]
     & &  &$\sigma_{ijk}^\ast$& 0.951 & 0.954 & 0.951 & 0.951 & 0.951 & 0.951\\[1.0ex] \cline{3-10}
     & & 100 &$\hat\sigma_{ijk}$& 0.876 & 0.854 & 0.876 & 0.863 & 0.830 & 0.863  \\ [1.0ex]
     & &  &$\sigma_{ijk}^\ast$& 0.950 & 0.943 & 0.950 & 0.950 & 0.934 & 0.950\\[1.0ex]\cline{2-10}
     & 500 & 30 &$\hat\sigma_{ijk}$ & 0.943 & 0.933 & 0.943 & 0.942 & 0.937 & 0.942  \\ [1.0ex]
     & &  &$\sigma_{ijk}^\ast$& 0.951 & 0.947 & 0.951 & 0.950 & 0.949 & 0.950\\[1.0ex]\cline{3-10}
     & & 100 &$\hat\sigma_{ijk}$& 0.925 & 0.918 & 0.925 & 0.921 & 0.912 & 0.921  \\ [1.0ex]
     & &  &$\sigma_{ijk}^\ast$& 0.950 & 0.948 & 0.950 & 0.951 & 0.950 & 0.951\\[1.0ex]\hline
     Hub & 200 & 30 & $\hat\sigma_{ijk}$& 0.938 & 0.928 & 0.938 & 0.927 & 0.905 & 0.927  \\ [1.0ex]
     & &  &$\sigma_{ijk}^\ast$& 0.950 & 0.949 & 0.950 & 0.950 & 0.947 & 0.950\\[1.0ex] \cline{3-10}
     & & 100 &$\hat\sigma_{ijk}$& 0.876 & 0.866 & 0.876 & 0.863 & 0.844 & 0.863  \\ [1.0ex]
     & &  &$\sigma_{ijk}^\ast$& 0.951 & 0.948 & 0.951 & 0.951 & 0.944 & 0.951\\[1.0ex]\cline{2-10}
     & 500 & 30 &$\hat\sigma_{ijk}$ & 0.943 & 0.940 & 0.943 & 0.942 & 0.938 & 0.942  \\ [1.0ex]
     & &  &$\sigma_{ijk}^\ast$& 0.951 & 0.949 & 0.951 & 0.951 & 0.949 & 0.951\\[1.0ex]\cline{3-10}
     & & 100 &$\hat\sigma_{ijk}$& 0.925 & 0.923 & 0.925 & 0.921 & 0.912 & 0.921  \\ [1.0ex]
     & &  &$\sigma_{ijk}^\ast$& 0.950 & 0.953 & 0.950 & 0.951 & 0.952 & 0.951\\[1.0ex]
  \hline \hline
\end{tabular}
}
\caption{\rkcr{Average coverage probabilities of the 95\% confidence intervals with the variance of $W_{l,jk}$ estimated by \eqref{eq:var_est}, referred to as $\hat\sigma_{ijk}$, and with the true variance of $W_{l,jk}$, referred to as $\sigma_{ijk}^\ast$.
Columns $\bB_l^\ast$, $\mathcal{S}(\bB_l^\ast)$ and $\mathcal{S}^c(\bB_l^\ast)$ show average coverage probabilities of off-diagonal parameters, non-zero off-diagonal parameters and zero off-diagonal parameters in all $\bB_l^\ast$'s, respectively.}} 
\label{tb:coverage}
\end{table}

Lastly, we evaluate the efficacy of the statistical inference procedure from Section \ref{sec:inference}. 
In Figure \ref{fig:ciplot}, we plot 95\% confidence intervals for entries in $\bB_1^\ast$ for one data replicate from the MA(1) model under Setting 2 with $q=100$, $n=500$. 
Specifically, the first 49 confidence intervals (indices 1–49) in Figure \ref{fig:ciplot} are shown for $B_{1,jk}^\ast$ for $|j-k|=1$ and the other 48 confidence intervals (indices 50–97) are shown for $|j-k|=2$. True parameter values are $B_{1,jk}^\ast=0.5$ for $|j-k|=1$ and $B_{1,jk}^\ast=0$ for $|j-k|=2$. It is seen that the 95\% confidence intervals show a good coverage. 
Finally, we evaluate the coverage probabilities for off-diagonal parameters in all $\bB_l^{\ast}$'s, $\mathcal{S}(\bB_l^\ast)$'s and $\mathcal{S}^c(\bB_l^\ast)$'s, respectively, in Table \ref{tb:coverage}.
Under each $n,q$ setting, we report the average coverage probabilities when the confidence intervals are calculated using the true variances of $W_{l,jk}$'s and empirical variances estimated using \eqref{eq:var_est}.
It is seen that the intervals calculated with empirical variances achieve a satisfactory coverage, and it approaches 95\% as the sample size increases.

\section{Real Data Analysis}
\label{sec:real}
We apply our proposed method \texttt{SparseCovReg} to the REMBRANDT study (GSE108476) that collects data on 178 patients with glioblastoma multiforme (GBM), the most common malignant form of brain tumor in adults and one of the most lethal of all cancers \citep{akhavan2010mtor}.
These 178 patients had undergone microarray and single-nucleotide polymorphism (SNP) chip profiling, with both gene expression and SNP data available for analysis. 
The raw data were pre-processed and normalized using standard pipelines; see \citet{gusev2018rembrandt} for more details. 
The main objectives of our analysis are to identify co-expression QTLs and recover both the population-level and individual-level covariance matrices of gene expressions.


For response variables, we consider the expression levels of 73 genes that belong to the human glioma pathway in the Kyoto Encyclopedia of Genes and Genomes (KEGG) database \citep{kanehisa2000kegg}. 
As covariates, we consider local SNPs (i.e., SNPs that fall within 2kb upstream and 0.5kb downstream of the gene) residing near those 73 genes, resulting in a total of 118 SNPs. SNPs are coded with “0” indicating homozygous in the major allele and “1” otherwise. Our analysis also includes age (continuous) and sex as covariates, bringing a total of 120 covariates and 326,821 parameters in the model \eqref{eq:model1}. 
Tuning parameters have been selected by 5-fold cross validation.

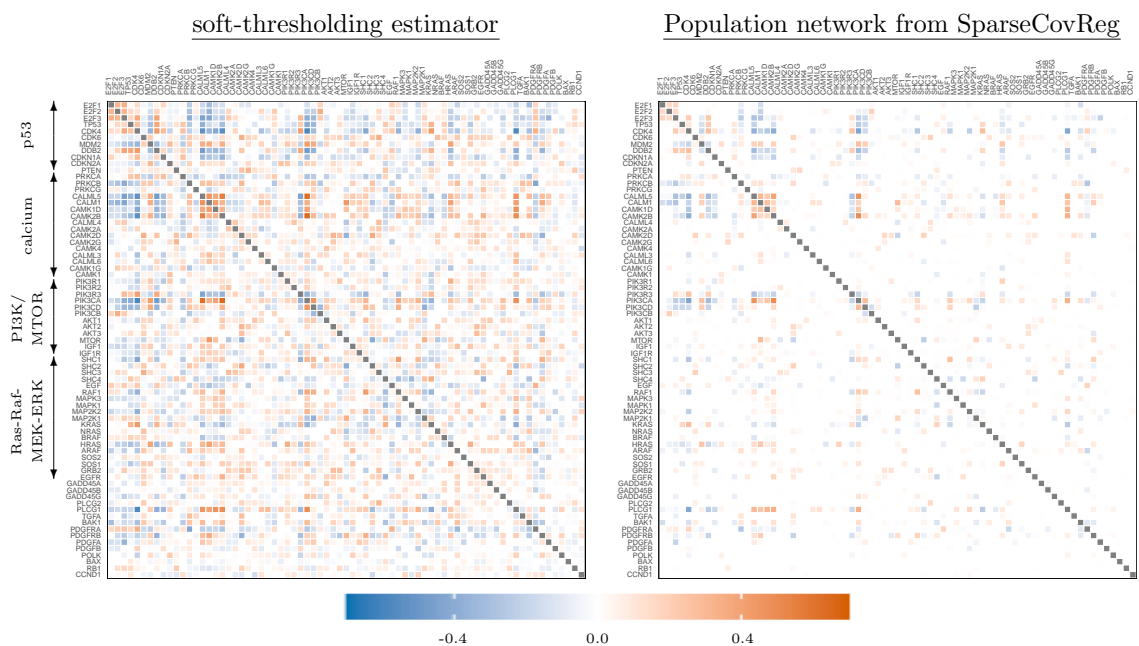
\begin{figure}
\centering
\scalebox{0.9}{
\mbox{
			\begin{overpic}[width=3.1in,angle=0]
				{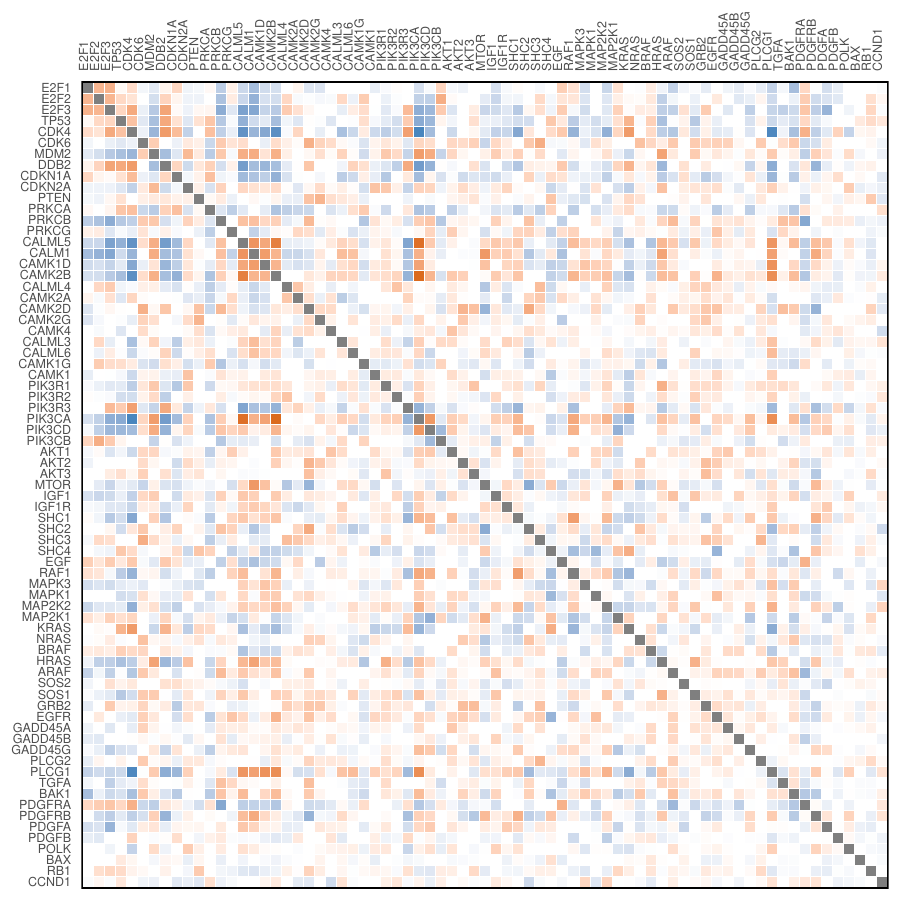}
            \put(25,104){\uline{\small{soft-thresholding estimator}}}

\put(-1,84.5){\vector(0,4){6.5}}
\put(-1,84.5){\vector(0,-4){6.5}}
\put(-7,82){\rotatebox{90}{\tiny{p53}}}
\put(-1,68){\vector(0,4){9.5}}
\put(-1,68){\vector(0,-4){10.0}}
\put(-7,62){\rotatebox{90}{\tiny{calcium}}}
\put(-1,50.5){\vector(0,4){7.0}}
\put(-1,50.5){\vector(0,-4){7.0}}
\put(-9,45.5){\rotatebox{90}{\tiny{PI3K/}}}
\put(-5,45){\rotatebox{90}{\tiny{MTOR}}}
\put(-1,31.5){\vector(0,4){11.5}}
\put(-1,31.5){\vector(0,-4){11.5}}
\put(-9,24.5){\rotatebox{90}{\tiny{Ras-Raf-}}}
\put(-5,23){\rotatebox{90}{\tiny{MEK-ERK}}}
            
			\end{overpic}
		
	   }
\hspace{-2mm}
\mbox{
			\begin{overpic}[width=3.1in,angle=0]
				{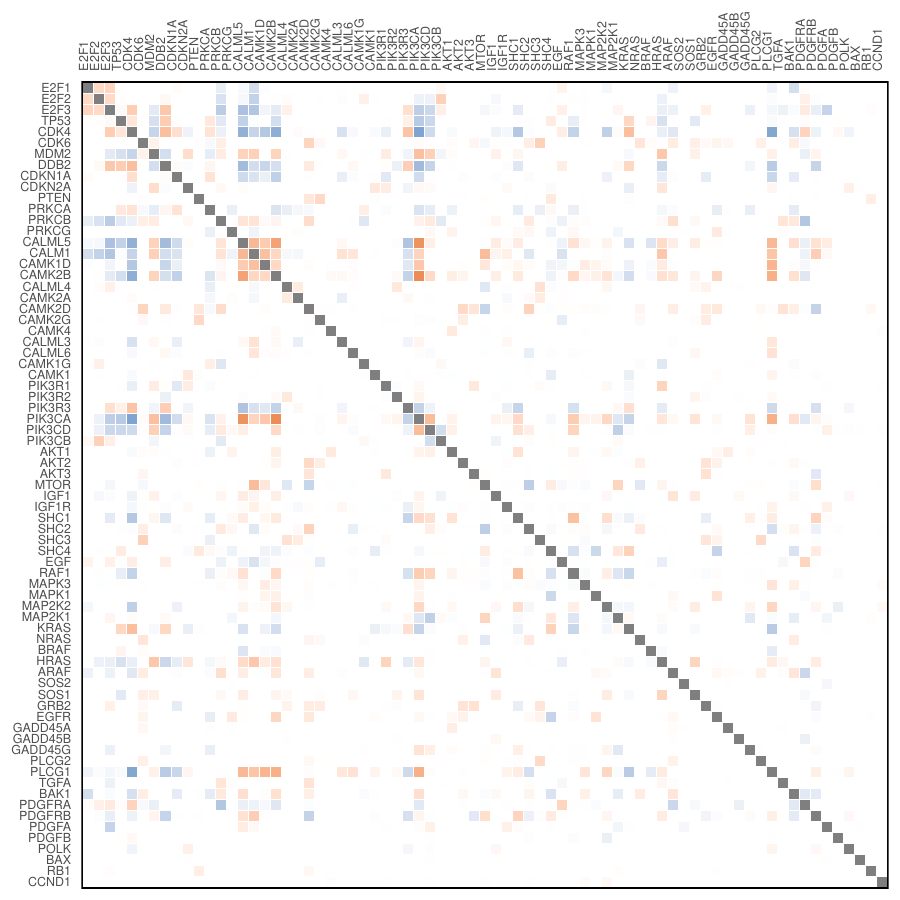}
            \put(10,104){\uline{\small{Population network from SparseCovReg}}}
			\end{overpic}
		
	   }
    }
\scalebox{0.9}{
\mbox{
			\begin{overpic}[width=3.1in, height=0.2in, angle=0]
				{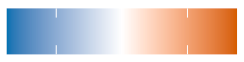}
			\end{overpic}
            \put(-56,-8){\tiny{0.4}}
            \put(-117,-8){\tiny{0.0}}
            \put(-179,-8){\tiny{-0.4}}
	   }
}
\vspace*{3mm}
\caption{Heatmaps of the population-level covariance estimates from \texttt{SparseSample} (left) and \texttt{SparseCovReg} (right). Positive values are shown in red and negative values are shown in blue.}
\label{fig:heatmap1}
\end{figure}

We first investigate the population-level co-expression matrix. 
In Figure \ref{fig:heatmap1}, we compare the soft-thresholding covariance estimator \citep{rothman2009generalized} with the population-level covariance $\bB_0$ obtained from \texttt{SparseCovReg}. 
It is seen that the soft-thresholding estimator shares some common patterns with \texttt{SparseCovReg} but is considerably more noisy. 
Using the population covariance matrix from \texttt{SparseCovReg}, we can identify high correlations between PIK3CA and genes in the calcium signaling pathway including CALML5, CALM1, CAMK1D and CAMK2B. This is reasonable as mutations in PIK3CA have been reported in multiple tumor types and PIK3CA is part of the PI3K/AKT/MTOR signaling pathway, one of the core pathways in human GBM \citep{cancer2008comprehensive}. The calcium signaling pathway also plays diverse roles in the progression of brain cancers \citep{maklad2019calcium}. 
Figure \ref{fig:heatmap1} shows negative correlations between PIK3CA and genes in the p53 signaling pathway, another core pathway in human GBM \citep{cancer2008comprehensive} and a potential target for inhibition in GBM treatments \citep{schroder2015cdk4, yin2021preclinical}.

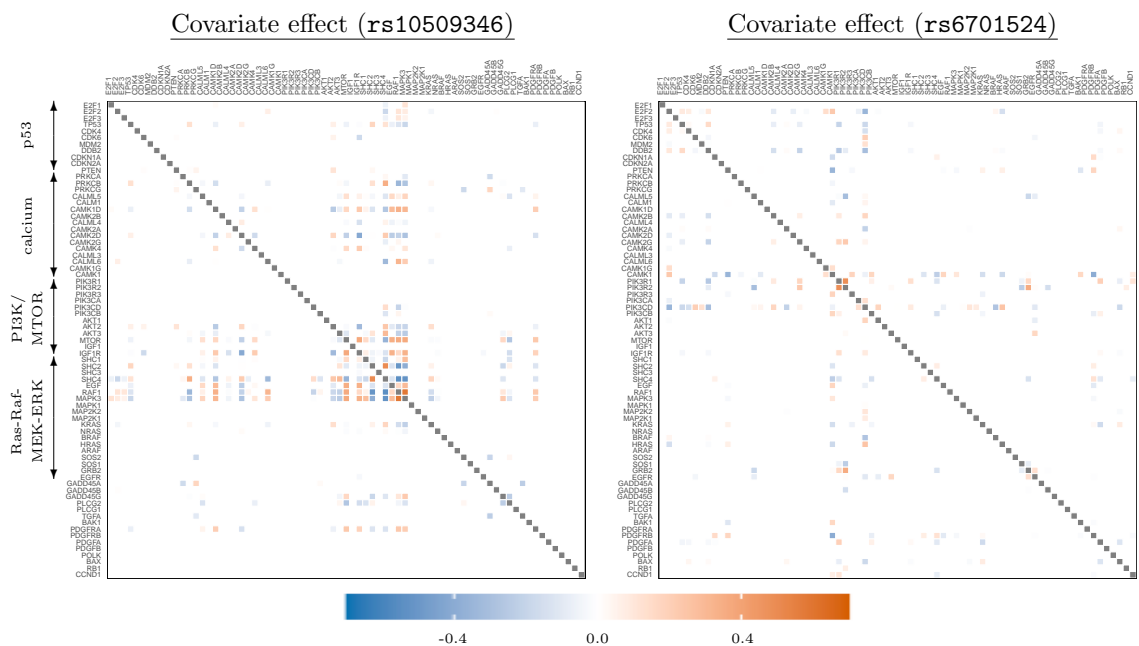
\begin{figure}
\centering
\scalebox{0.9}{
\mbox{
			\begin{overpic}[width=3.1in,angle=0]
				{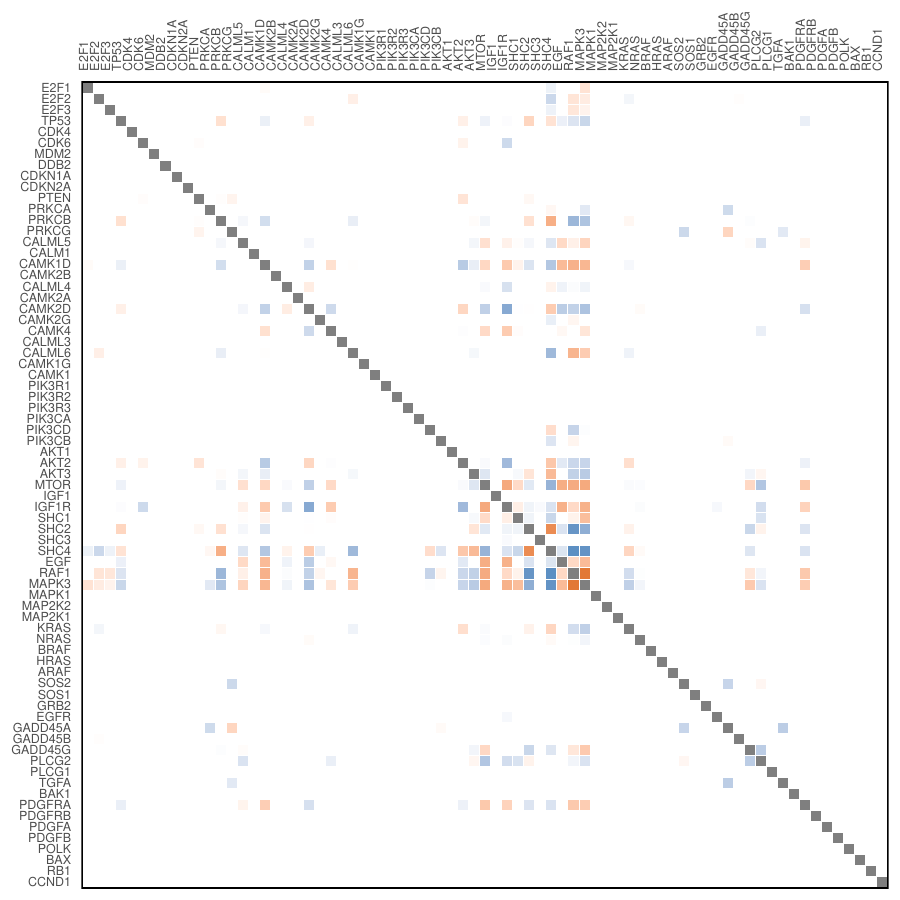}
            \put(21,104){\uline{\small{Covariate effect (\texttt{rs10509346})}}}

\put(-1,84.5){\vector(0,4){6.5}}
\put(-1,84.5){\vector(0,-4){6.5}}
\put(-7,82){\rotatebox{90}{\tiny{p53}}}
\put(-1,68){\vector(0,4){9.5}}
\put(-1,68){\vector(0,-4){10.0}}
\put(-7,62){\rotatebox{90}{\tiny{calcium}}}
\put(-1,50.5){\vector(0,4){7.0}}
\put(-1,50.5){\vector(0,-4){7.0}}
\put(-9,45.5){\rotatebox{90}{\tiny{PI3K/}}}
\put(-5,45){\rotatebox{90}{\tiny{MTOR}}}
\put(-1,31.5){\vector(0,4){11.5}}
\put(-1,31.5){\vector(0,-4){11.5}}
\put(-9,24.5){\rotatebox{90}{\tiny{Ras-Raf-}}}
\put(-5,23){\rotatebox{90}{\tiny{MEK-ERK}}}
            
			\end{overpic}
		
	   }
\hspace{-2mm}
\mbox{
			\begin{overpic}[width=3.1in,angle=0]
				{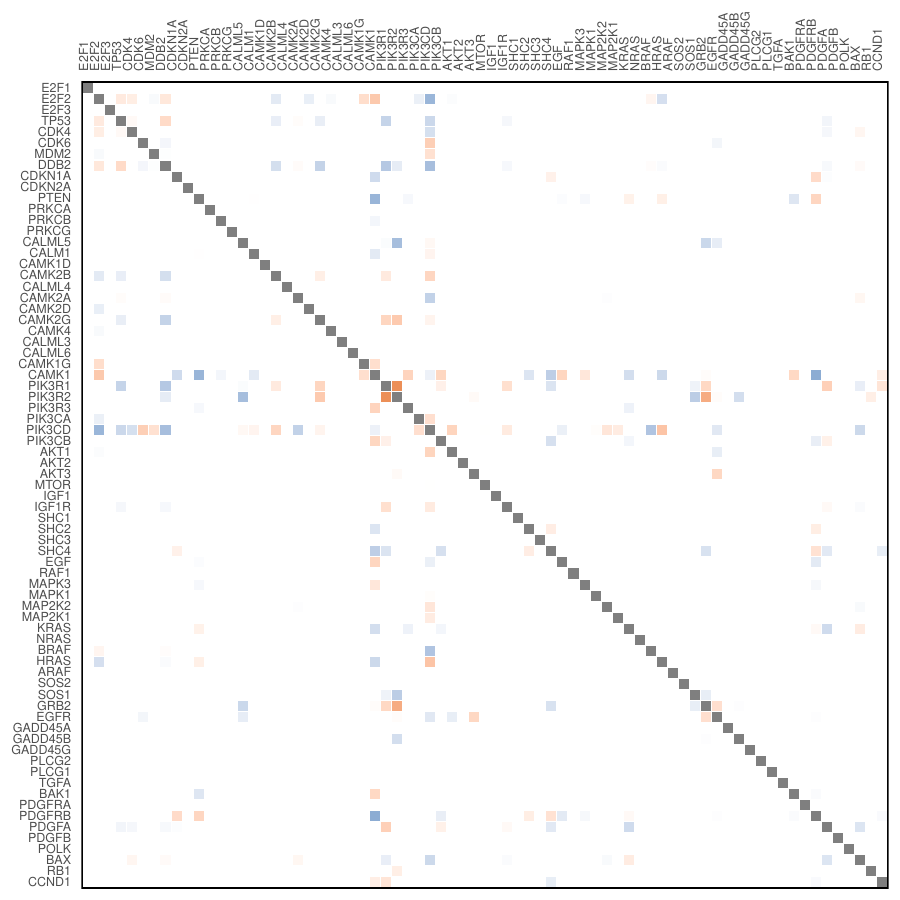}
            \put(22,104){\uline{\small{Covariate effect (\texttt{rs6701524})}}}
			\end{overpic}
		
	   }
    }
\scalebox{0.9}{
\mbox{
			\begin{overpic}[width=3.1in, height=0.2in, angle=0]
				{figures/heat_legend1}
			\end{overpic}
		    \put(-56,-8){\tiny{0.4}}
            \put(-117,-8){\tiny{0.0}}
            \put(-179,-8){\tiny{-0.4}}
	   }
}
\vspace*{3mm}
\caption{Heatmaps of identified nonzero covariate effects. Positive values are shown in red and negative values are shown in blue.}
\label{fig:heatmap2}
\end{figure}

Next, we examine the covariate effects on the covariance matrix. Non-zero effects have been identified for six SNPs: \texttt{rs6701524}, \texttt{rs10509346}, \texttt{rs10519201}, \texttt{rs1347069}, \texttt{rs503314}, and \texttt{rs306098}. 
The non-zero effects of \texttt{rs10509346} and \texttt{rs6701524} are shown in Figure \ref{fig:heatmap2} and their network effects after the debiased inference procedure are shown in Figure \ref{fig:netmap}. Results for the other four SNPs are included in Section S8 of the Supplementary Materials. 
Interestingly, these covariate effects are not easily observable from the soft-thresholding estimator in Figure \ref{fig:heatmap1}, suggesting that, by fitting model (\ref{eq:model1}), we may find some covariate-modulated co-expression patterns that can otherwise be overlooked.

From the left plot of Figure \ref{fig:netmap}, it is seen that \texttt{rs10509346}, residing in CAMK2G, notably affect co-expressions among genes in the Ras-Raf-MEK-ERK signaling pathway including EGF, SHC4, RAF1 and MAPK3. Also, their co-expressions with CAMK2D and CALML5 in the calcium signaling pathway are affected by \texttt{rs10509346}. This agrees with the findings that the Ras-Raf-MEK-ERK pathway is modulated by Ca$^{+2}$ and calmodulin \citep{agell2002modulation, zhang2022high}. Furthermore, this SNP is found to affect the co-expressions of MTOR, part of the PI3K/AKT/MTOR pathway, with genes in the Ras-Raf-MEK-ERK pathway. This result is interesting because MTOR is a key mediator of PI3K/AKT/MTOR signaling, and is known to cooperate with alterations in other signaling pathways that are also commonly activated in GBM patients, such as the Ras-Raf-MEK-ERK pathway \citep{akhavan2010mtor}.

The right plot of Figure \ref{fig:netmap}, shows \texttt{rs6701524}, residing in MTOR, affects co-expressions of genes in the PI3K/MTOR pathway. In particular, co-expressions of PIK3CD (and PIK3CB) with other genes are affected by this SNP.
This is an interesting finding as PI3K/MTOR is a key pathway in the development and progression of GBM, and the inhibition of PI3K/MTOR signaling was found effective in increasing survival with GBM tumor \citep{batsios2019pi3k}.
Co-expressions affected by other SNPs are also worth noting. For example, \texttt{rs306098} has been found to affect co-expressions of SHC2 with CDK4/6 and MTOR, which is interesting because the combination of CDK4/6 and MTOR inhibition has been investigated as a potential therapeutic strategy in GBM \citep{olmez2017combined}.

\rkcr{Lastly, we validate the model by assessing the stability of the selected SNPs. Specifically, we randomly split the data into equal-sized training and testing sets 100 times. In each iteration, we fit \texttt{SparseCovReg} on both the training and testing data and record the effective SNPs with a nonzero coefficient matrix. The average number of effective SNPs selected over 100 iterations was 11.6 for the training data and 11.7 for the test data. Among these, seven SNPs were selected in both the training and testing sets more than 10 times: \texttt{rs6701524}, \texttt{rs10509346}, \texttt{rs10519201}, \texttt{rs1347069}, \texttt{rs503314}, \texttt{rs306098}, and \texttt{rs2053158}. Notably, this set includes all six SNPs originally selected by \texttt{SparseCovReg} when fitted to the full dataset. In particular, \texttt{rs6701524} was selected in both the training and testing sets 66 times.}


\begin{figure}
\centering
\scalebox{1.0}{
\mbox{
			\begin{overpic}[width=3.1in,angle=0]
				{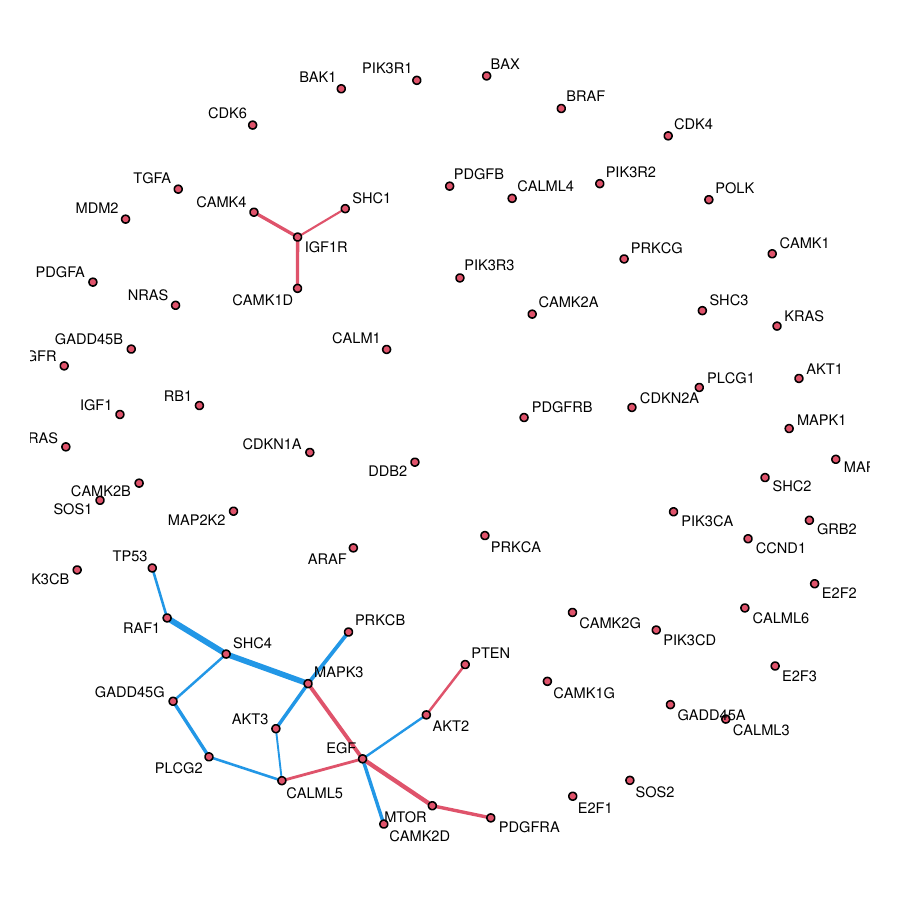}
            \put(37,104){{\uline{\small{\texttt{rs10509346}}}}}
			\end{overpic}
		
	   }
\hspace{-2mm}
\mbox{
			\begin{overpic}[width=3.1in,angle=0]
				{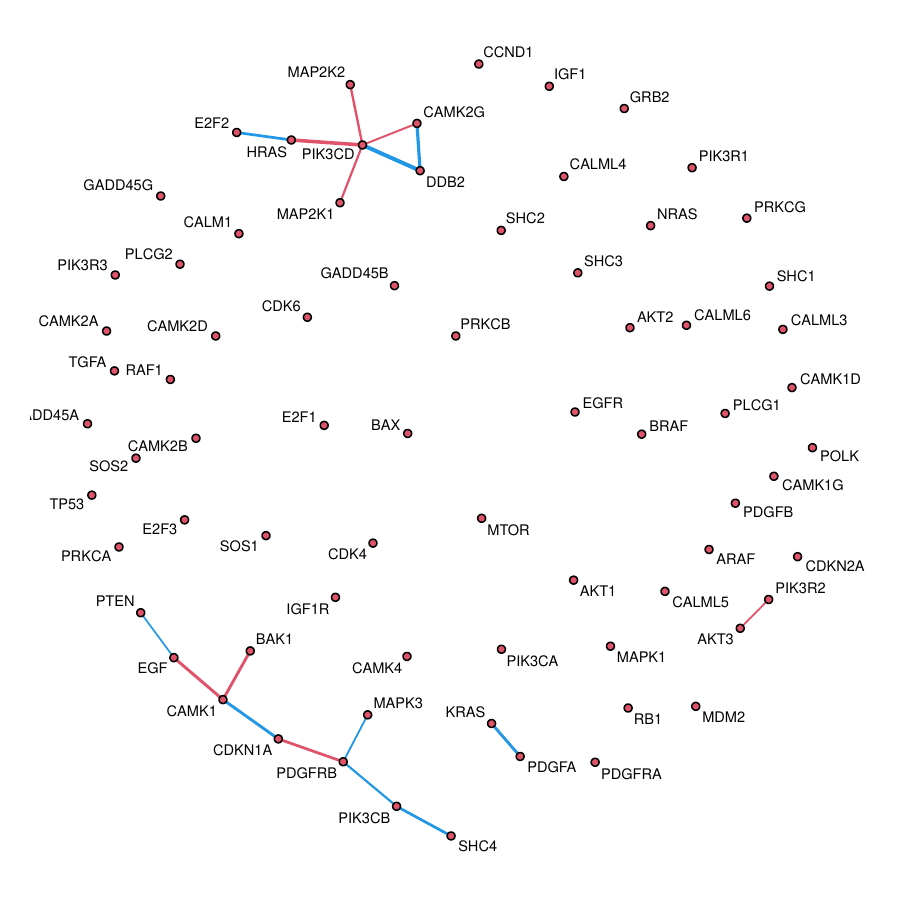}
            \put(38,104){{\uline{\small{\texttt{rs6701524}}}}}
			\end{overpic}
		
	   }
    }

\vspace*{6mm}

\caption{Network effects of \texttt{rs10509346} (left) and \texttt{rs6701524} (right) identified using the debiased inference procedure in Section \ref{sec:inference} with significance level $\alpha=1\%$ and Bonferroni correction $1-\alpha/\{p(p-1)/2\}$. Positive effects are shown in red and negative effects are shown in blue. Edge weights in the network graphs are proportional to the effect size.}
\label{fig:netmap}
\end{figure}

\section{Discussion}
\label{sec:conc}

\rkcr{Our current approach does not take into account the correlation between $z_{ij}z_{ik}$ and $z_{ij'}z_{ik'}$ in \eqref{eq:model10} although they are likely to be correlated, for example, when $j=j'$. When estimating $\bB_h$'s, ignoring such correlations does not affect the consistency of our estimators, as established by Theorems \ref{thm2} and \ref{thm4}. It also does not impact the validity of our inferential procedure in Section \ref{sec:inference}, since we debias $\wh{\bB}_{\mydot,jk}$ for each $j,k \in [p]$ separately, eliminating the need to consider the correlation between $z_{ij}z_{ik}$ and $z_{ij'}z_{ik'}$. 
However, our least squares estimators may not be as efficient as the generalized least squares estimator. Computing the generalized least squares estimator requires consistently estimating $\text{Cov}(z_{ij}z_{ik}, z_{ij'}z_{ik'})$ for all $j,k,j',k'$, which can be computationally prohibitive due to the large number of parameters. A potential solution is to adopt a penalized generalized estimating equations (GEE) approach \citep{fu2003penalized}. This would require redesigning our estimation and inferential procedure. In particular, conducting debiased inference under this setting is nontrivial, and we leave it for future research.}

\rkcr{When a new observation is available, it may be possible that the new observation $\bx^{(new)}$ has covariate values outside $[u_l, v_l]$. In such cases, one only needs to adjust $\delta$ in \eqref{eq:pd_approx} to ensure that \eqref{eq:pd_condition} holds. The unconstrained minimizer of \eqref{eq:obj_ftn2} does not need to be re-calculated, as 
it is unaffected by the range of covariates.}
\rkcr{Some covariates may be specific to the response variables rather than the individuals. For example, there could be gene-specific variables when constructing gene networks. 
Model \eqref{eq:model1} can be modified to account for such variable-specific covariates. Specifically, given $G$ variable-specific covariates, we can modify model \eqref{eq:model1} by replacing the term $\bB_0$ with the covariance regression model in \citet{zou2017covariance} as below:
\begin{equation*} 
\bSigma(\bx) = \sum_{g=1}^G \beta_g \bD_g + \sum_{l=1}^{q} x_l \bB_l,
\end{equation*}
where $\bD_g$, $g \in [G]$ is a known similarity matrix of the variable-specific covariates. 
To estimate all parameters in the above model, we may replace Step 1 in Algorithm \ref{alg1} with the estimation methods proposed in \citet{zou2017covariance} for estimating $\beta_1,\ldots,\beta_G$.}

\rkcr{We did not consider any a priori structural information in the covariance matrix. If any structural information is available, our method can be modified or extended to account for such information. For example, if the response variables are from two communities, it may be reasonable to assume that the covariance entries within each community take similar values. 
Such similarity within each community can be taken into account by adding the fused penalty \citep{tibshirani2005sparsity}, which enforces similarity among coefficients.
This requires different model assumptions and optimization procedure, and we leave it for future research.} 
\rkcr{Currently, our method enforces sparsity in the coefficient matrices $\bB_0,\bB_1,\ldots,\bB_q$, rather than directly in the covariance matrix $\bSigma(x)$, allowing for greater flexibility in modeling different covariance structures across subgroups of subjects. When it is of interest to enforce sparsity directly in $\bSigma(x)$, the penalty function can be modified to incorporate a hierarchical structure. Specifically, we can assume that covariate-specific effects $B_{l,jk}, l\in[q]$ are non-zero only when the population-level effect $B_{0,jk}$ is non-zero, leading to the desired sparsity structure. Our sparse covariance regression framework can be extended to accommodate such a hierarchical structure in the penalty term \citep{tibshirani2020pliable, kim2021svreg}.}


Next, we comment on the model interpretability after scaling the covariates to be in $[-1,1]$ as in Assumption \ref{ass:covariate_dist}. Given bounded covariates $x_l \in [u_l,v_l]$ for all $l \in [q]$, 
consider the covariance regression model 
\bse
\bSigma(\bx) = \bB_0 + \sum_{l=1}^{q} \frac{x_l - (v_l+u_l)/2}{(v_l-u_l)/2} \bB_l.
\ese
The above equation can be rewritten as 
\bse
\bSigma(\bx) = \bm\bar{\bB}_0 + \sum_{l=1}^{q} x_l \bm\bar{\bB}_l.
\ese
where $\bm\bar{\bB}_0 = \bB_0 - \sum_{l=1}^{q} (v_l+u_l)/(v_l-u_l) \bB_l$ and $\bm\bar{\bB}_l = \bB_l/\{(v_l-u_l)/2\}$. Note that $\bm\bar{\bB}_l$ and $\bB_l$ only differ by a positive scalar, and they share the same sparsity pattern. Hence, parameter estimates can be interpreted with covariates before the transformation.

\bibliographystyle{asa}
\begingroup
\baselineskip=17.5pt
\bibliography{covlm}

\begin{thebibliography}{58}
\newcommand{\enquote}[1]{``#1''}
\expandafter\ifx\csname natexlab\endcsname\relax\def\natexlab#1{#1}\fi

\bibitem[{Agell et~al.(2002)Agell, Bachs, Rocamora, and
  Villalonga}]{agell2002modulation}
Agell, N., Bachs, O., Rocamora, N., and Villalonga, P. (2002),
  \enquote{Modulation of the Ras/Raf/MEK/ERK pathway by Ca2+, and calmodulin,}
  \textit{Cellular signalling}, 14, 649--654.

\bibitem[{Akhavan et~al.(2010)Akhavan, Cloughesy, and
  Mischel}]{akhavan2010mtor}
Akhavan, D., Cloughesy, T.~F., and Mischel, P.~S. (2010), \enquote{mTOR
  signaling in glioblastoma: lessons learned from bench to bedside,}
  \textit{Neuro-oncology}, 12, 882--889.

\bibitem[{Alakus et~al.(2023)Alakus, Larocque, and
  Labbe}]{alakus2022covariance}
Alakus, C., Larocque, D., and Labbe, A. (2023), \enquote{Covariance regression
  with random forests,} \textit{BMC bioinformatics}, 24, 258.

\bibitem[{Anderson(1973)}]{anderson1973asymptotically}
Anderson, T.~W. (1973), \enquote{Asymptotically efficient estimation of
  covariance matrices with linear structure,} \textit{The Annals of
  Statistics}, 1, 135--141.

\bibitem[{Argyriou et~al.(2008)Argyriou, Evgeniou, and
  Pontil}]{argyriou2008convex}
Argyriou, A., Evgeniou, T., and Pontil, M. (2008), \enquote{Convex multi-task
  feature learning,} \textit{Machine learning}, 73, 243--272.

\bibitem[{Batsios et~al.(2019)Batsios, Viswanath, Subramani, Najac, Gillespie,
  Santos, Molloy, Pieper, and Ronen}]{batsios2019pi3k}
Batsios, G., Viswanath, P., Subramani, E., Najac, C., Gillespie, A.~M., Santos,
  R.~D., Molloy, A.~R., Pieper, R.~O., and Ronen, S.~M. (2019),
  \enquote{PI3K/mTOR inhibition of IDH1 mutant glioma leads to reduced 2HG
  production that is associated with increased survival,} \textit{Scientific
  reports}, 9, 10521.

\bibitem[{Bickel et~al.(2008{\natexlab{a}})Bickel, Levina,
  et~al.}]{bickel2008covariance}
Bickel, P.~J., Levina, E., et~al. (2008{\natexlab{a}}), \enquote{Covariance
  regularization by thresholding,} \textit{The Annals of Statistics}, 36,
  2577--2604.

\bibitem[{Bickel et~al.(2008{\natexlab{b}})Bickel, Levina,
  et~al.}]{bickel2008regularized}
--- (2008{\natexlab{b}}), \enquote{Regularized estimation of large covariance
  matrices,} \textit{The Annals of Statistics}, 36, 199--227.

\bibitem[{Bien and Tibshirani(2011)}]{bien2011sparse}
Bien, J. and Tibshirani, R.~J. (2011), \enquote{Sparse estimation of a
  covariance matrix,} \textit{Biometrika}, 98, 807--820.

\bibitem[{B{\"u}hlmann and Van~de Geer(2015)}]{buhlmann2015high}
B{\"u}hlmann, P. and Van~de Geer, S. (2015), \enquote{High-dimensional
  inference in misspecified linear models,} \textit{Electronic Journal of
  Statistics}, 9, 1449--1473.

\bibitem[{Butte et~al.(2000)Butte, Tamayo, Slonim, Golub, and
  Kohane}]{butte2000discovering}
Butte, A.~J., Tamayo, P., Slonim, D., Golub, T.~R., and Kohane, I.~S. (2000),
  \enquote{Discovering functional relationships between RNA expression and
  chemotherapeutic susceptibility using relevance networks,}
  \textit{Proceedings of the National Academy of Sciences}, 97, 12182--12186.

\bibitem[{Cai et~al.(2022)Cai, Zhang, and Zhou}]{cai2022sparse}
Cai, T.~T., Zhang, A.~R., and Zhou, Y. (2022), \enquote{Sparse group lasso:
  Optimal sample complexity, convergence rate, and statistical inference,}
  \textit{IEEE Transactions on Information Theory}, 68, 5975--6002.

\bibitem[{Chiu et~al.(1996)Chiu, Leonard, and Tsui}]{chiu1996matrix}
Chiu, T.~Y., Leonard, T., and Tsui, K.-W. (1996), \enquote{The
  matrix-logarithmic covariance model,} \textit{Journal of the American
  Statistical Association}, 91, 198--210.

\bibitem[{El~Karoui et~al.(2010)}]{el2010high}
El~Karoui, N. et~al. (2010), \enquote{High-dimensionality effects in the
  Markowitz problem and other quadratic programs with linear constraints: Risk
  underestimation,} \textit{The Annals of Statistics}, 38, 3487--3566.

\bibitem[{Fox and Dunson(2015)}]{fox2015bayesian}
Fox, E.~B. and Dunson, D.~B. (2015), \enquote{Bayesian nonparametric covariance
  regression,} \textit{The Journal of Machine Learning Research}, 16,
  2501--2542.

\bibitem[{Franks(2021)}]{franks2021reducing}
Franks, A.~M. (2021), \enquote{Reducing subspace models for large-scale
  covariance regression,} \textit{Biometrics}.

\bibitem[{Fu(2003)}]{fu2003penalized}
Fu, W.~J. (2003), \enquote{Penalized estimating equations,}
  \textit{Biometrics}, 59, 126--132.

\bibitem[{Gardner et~al.(2003)Gardner, Di~Bernardo, Lorenz, and
  Collins}]{gardner2003inferring}
Gardner, T.~S., Di~Bernardo, D., Lorenz, D., and Collins, J.~J. (2003),
  \enquote{Inferring genetic networks and identifying compound mode of action
  via expression profiling,} \textit{Science}, 301, 102--105.

\bibitem[{Gusev et~al.(2018)Gusev, Bhuvaneshwar, Song, Zenklusen, Fine, and
  Madhavan}]{gusev2018rembrandt}
Gusev, Y., Bhuvaneshwar, K., Song, L., Zenklusen, J.-C., Fine, H., and
  Madhavan, S. (2018), \enquote{The REMBRANDT study, a large collection of
  genomic data from brain cancer patients,} \textit{Scientific data}, 5, 1--9.

\bibitem[{Hoff and Niu(2012)}]{hoff2012covariance}
Hoff, P.~D. and Niu, X. (2012), \enquote{A covariance regression model,}
  \textit{Statistica Sinica}, 729--753.

\bibitem[{Huang et~al.(2006)Huang, Liu, Pourahmadi, and
  Liu}]{huang2006covariance}
Huang, J.~Z., Liu, N., Pourahmadi, M., and Liu, L. (2006), \enquote{Covariance
  matrix selection and estimation via penalised normal likelihood,}
  \textit{Biometrika}, 93, 85--98.

\bibitem[{Javanmard and Montanari(2014)}]{javanmard2014confidence}
Javanmard, A. and Montanari, A. (2014), \enquote{Confidence intervals and
  hypothesis testing for high-dimensional regression,} \textit{The Journal of
  Machine Learning Research}, 15, 2869--2909.

\bibitem[{Kanehisa and Goto(2000)}]{kanehisa2000kegg}
Kanehisa, M. and Goto, S. (2000), \enquote{KEGG: kyoto encyclopedia of genes
  and genomes,} \textit{Nucleic acids research}, 28, 27--30.

\bibitem[{Kim et~al.(2021)Kim, Mueller, and Garcia}]{kim2021svreg}
Kim, R., Mueller, S., and Garcia, T.~P. (2021), \enquote{svReg: Structural
  varying-coefficient regression to differentiate how regional brain atrophy
  affects motor impairment for Huntington disease severity groups,}
  \textit{Biometrical Journal}, 63, 1254--1271.

\bibitem[{Lam and Fan(2009)}]{lam2009sparsistency}
Lam, C. and Fan, J. (2009), \enquote{Sparsistency and rates of convergence in
  large covariance matrix estimation,} \textit{Annals of statistics}, 37, 4254.

\bibitem[{Langfelder and Horvath(2008)}]{langfelder2008wgcna}
Langfelder, P. and Horvath, S. (2008), \enquote{WGCNA: an R package for
  weighted correlation network analysis,} \textit{BMC bioinformatics}, 9,
  1--13.

\bibitem[{Ledoit and Wolf(2004)}]{ledoit2004well}
Ledoit, O. and Wolf, M. (2004), \enquote{A well-conditioned estimator for
  large-dimensional covariance matrices,} \textit{Journal of multivariate
  analysis}, 88, 365--411.

\bibitem[{Li et~al.(2015)Li, Nan, and Zhu}]{li2015multivariate}
Li, Y., Nan, B., and Zhu, J. (2015), \enquote{Multivariate sparse group lasso
  for the multivariate multiple linear regression with an arbitrary group
  structure,} \textit{Biometrics}, 71, 354--363.

\bibitem[{Li et~al.(2010)Li, Wang, and Carroll}]{li2010generalized}
Li, Y., Wang, N., and Carroll, R.~J. (2010), \enquote{Generalized functional
  linear models with semiparametric single-index interactions,} \textit{Journal
  of the American Statistical Association}, 105, 621--633.

\bibitem[{Maklad et~al.(2019)Maklad, Sharma, and Azimi}]{maklad2019calcium}
Maklad, A., Sharma, A., and Azimi, I. (2019), \enquote{Calcium signaling in
  brain cancers: roles and therapeutic targeting,} \textit{Cancers}, 11, 145.

\bibitem[{Network et~al.(2008)}]{cancer2008comprehensive}
Network et~al. (2008), \enquote{Comprehensive genomic characterization defines
  human glioblastoma genes and core pathways,} \textit{Nature}, 455,
  1061--1068.

\bibitem[{Olmez et~al.(2017)Olmez, Brenneman, Xiao, Serbulea, Benamar, Zhang,
  Manigat, Abbas, Lee, Nakano, et~al.}]{olmez2017combined}
Olmez, I., Brenneman, B., Xiao, A., Serbulea, V., Benamar, M., Zhang, Y.,
  Manigat, L., Abbas, T., Lee, J., Nakano, I., et~al. (2017), \enquote{Combined
  CDK4/6 and mTOR inhibition is synergistic against glioblastoma via multiple
  mechanisms,} \textit{Clinical Cancer Research}, 23, 6958--6968.

\bibitem[{Park(2023)}]{park2023bayesian}
Park, H.~G. (2023), \enquote{Bayesian estimation of covariate assisted
  principal regression for brain functional connectivity,} \textit{arXiv
  preprint arXiv:2306.07181}.

\bibitem[{Pourahmadi(1999)}]{pourahmadi1999joint}
Pourahmadi, M. (1999), \enquote{Joint mean-covariance models with applications
  to longitudinal data: Unconstrained parameterisation,} \textit{Biometrika},
  86, 677--690.

\bibitem[{Qiu and Liyanage(2019)}]{qiu2019threshold}
Qiu, Y. and Liyanage, J.~S. (2019), \enquote{Threshold selection for covariance
  estimation,} \textit{Biometrics}, 75, 895--905.

\bibitem[{Rothman et~al.(2009)Rothman, Levina, and
  Zhu}]{rothman2009generalized}
Rothman, A.~J., Levina, E., and Zhu, J. (2009), \enquote{Generalized
  thresholding of large covariance matrices,} \textit{Journal of the American
  Statistical Association}, 104, 177--186.

\bibitem[{Schr{\"o}der and McDonald(2015)}]{schroder2015cdk4}
Schr{\"o}der, L.~B. and McDonald, K.~L. (2015), \enquote{CDK4/6 inhibitor
  PD0332991 in glioblastoma treatment: does it have a future?}
  \textit{Frontiers in oncology}, 5, 259.

\bibitem[{Simon et~al.(2013)Simon, Friedman, Hastie, and
  Tibshirani}]{Simon2013}
Simon, N., Friedman, J., Hastie, T., and Tibshirani, R. (2013), \enquote{A
  sparse-group lasso,} \textit{Journal of Computational and Graphical
  Statistics}, 22, 231--245.

\bibitem[{Su et~al.(2023)Su, Xu, Shan, Cai, Zhao, and Zhang}]{su2023cell}
Su, C., Xu, Z., Shan, X., Cai, B., Zhao, H., and Zhang, J. (2023),
  \enquote{Cell-type-specific co-expression inference from single cell
  RNA-sequencing data,} \textit{Nature Communications}, 14, 4846.

\bibitem[{Tibshirani and Friedman(2020)}]{tibshirani2020pliable}
Tibshirani, R. and Friedman, J. (2020), \enquote{A pliable lasso,}
  \textit{Journal of Computational and Graphical Statistics}, 29, 215--225.

\bibitem[{Tibshirani et~al.(2005)Tibshirani, Saunders, Rosset, Zhu, and
  Knight}]{tibshirani2005sparsity}
Tibshirani, R., Saunders, M., Rosset, S., Zhu, J., and Knight, K. (2005),
  \enquote{Sparsity and smoothness via the fused lasso,} \textit{Journal of the
  Royal Statistical Society Series B: Statistical Methodology}, 67, 91--108.

\bibitem[{Tseng(2001)}]{tseng2001convergence}
Tseng, P. (2001), \enquote{Convergence of a block coordinate descent method for
  nondifferentiable minimization,} \textit{Journal of optimization theory and
  applications}, 109, 475--494.

\bibitem[{Van Der~Wijst et~al.(2018)Van Der~Wijst, de~Vries, Brugge, Westra,
  and Franke}]{van2018integrative}
Van Der~Wijst, M.~G., de~Vries, D.~H., Brugge, H., Westra, H.-J., and Franke,
  L. (2018), \enquote{An integrative approach for building personalized gene
  regulatory networks for precision medicine,} \textit{Genome medicine}, 10,
  1--15.

\bibitem[{Vershynin(2018)}]{vershynin2018high}
Vershynin, R. (2018), \textit{High-dimensional probability: An introduction
  with applications in data science}, vol.~47, Cambridge university press.

\bibitem[{Vierstra et~al.(2020)Vierstra, Lazar, Sandstrom, Halow, Lee, Bates,
  Diegel, Dunn, Neri, Haugen, et~al.}]{vierstra2020global}
Vierstra, J., Lazar, J., Sandstrom, R., Halow, J., Lee, K., Bates, D., Diegel,
  M., Dunn, D., Neri, F., Haugen, E., et~al. (2020), \enquote{Global reference
  mapping of human transcription factor footprints,} \textit{Nature}, 583,
  729--736.

\bibitem[{Wu and Pourahmadi(2003)}]{wu2003nonparametric}
Wu, W.~B. and Pourahmadi, M. (2003), \enquote{Nonparametric estimation of large
  covariance matrices of longitudinal data,} \textit{Biometrika}, 90, 831--844.

\bibitem[{Xu and Lange(2022)}]{xu2022proximal}
Xu, J. and Lange, K. (2022), \enquote{A proximal distance algorithm for
  likelihood-based sparse covariance estimation,} \textit{Biometrika}, 109,
  1047--1066.

\bibitem[{Xue et~al.(2012)Xue, Ma, and Zou}]{xue2012positive}
Xue, L., Ma, S., and Zou, H. (2012), \enquote{Positive-definite
  $\ell$1-penalized estimation of large covariance matrices,} \textit{Journal
  of the American Statistical Association}, 107, 1480--1491.

\bibitem[{Yin et~al.(2021)Yin, Yao, Wang, Huang, Mazuranic, and
  Yin}]{yin2021preclinical}
Yin, L., Yao, Z., Wang, Y., Huang, J., Mazuranic, M., and Yin, A. (2021),
  \enquote{In Preclinical evaluation of novel CDK4/6 inhibitor GLR2007 in
  glioblastoma models,} \textit{J. Clin. Oncol}, 39, e14023.

\bibitem[{Yuan and Lin(2006)}]{Yuan2006}
Yuan, M. and Lin, Y. (2006), \enquote{Model selection and estimation in
  regression with grouped variables,} \textit{Journal of the Royal Statistical
  Society: Series B (Statistical Methodology)}, 68, 49--67.

\bibitem[{Zhang and Zhang(2014)}]{zhang2014confidence}
Zhang, C.-H. and Zhang, S.~S. (2014), \enquote{Confidence intervals for low
  dimensional parameters in high dimensional linear models,} \textit{Journal of
  the Royal Statistical Society: Series B: Statistical Methodology}, 217--242.

\bibitem[{Zhang and Li(2023)}]{zhang2022high}
Zhang, J. and Li, Y. (2023), \enquote{High-dimensional Gaussian graphical
  regression models with covariates,} \textit{Journal of the American
  Statistical Association}, 118, 2088--2100.

\bibitem[{Zhang et~al.(2020)Zhang, Sun, and Li}]{zhang2020mixed}
Zhang, J., Sun, W.~W., and Li, L. (2020), \enquote{Mixed-effect time-varying
  network model and application in brain connectivity analysis,}
  \textit{Journal of the American Statistical Association}, 115, 2022--2036.

\bibitem[{{}Zhang et~al.(2023){}Zhang, Sun, and Li}]{zhang2023generalized}
{}Zhang, J., Sun, W.~W., and Li, L. (2023), \enquote{Generalized connectivity
  matrix response regression with applications in brain connectivity studies,}
  \textit{Journal of Computational and Graphical Statistics}, 32, 252--262.

\bibitem[{Zhang and Zhao(2023)}]{zhang2023eqtl}
Zhang, J. and Zhao, H. (2023), \enquote{eQTL studies: from bulk tissues to
  single cells,} \textit{Journal of Genetics and Genomics}, 50, 925--933.

\bibitem[{Zhao et~al.(2021)Zhao, Wang, Mostofsky, Caffo, and
  Luo}]{zhao2021covariate}
Zhao, Y., Wang, B., Mostofsky, S.~H., Caffo, B.~S., and Luo, X. (2021),
  \enquote{Covariate assisted principal regression for covariance matrix
  outcomes,} \textit{Biostatistics}, 22, 629--645.

\bibitem[{Zou et~al.(2022)Zou, Lan, Li, and Tsai}]{zou2022inference}
Zou, T., Lan, W., Li, R., and Tsai, C.-L. (2022), \enquote{Inference on
  covariance-mean regression,} \textit{Journal of Econometrics}, 230, 318--338.

\bibitem[{Zou et~al.(2017)Zou, Lan, Wang, and Tsai}]{zou2017covariance}
Zou, T., Lan, W., Wang, H., and Tsai, C.-L. (2017), \enquote{Covariance
  regression analysis,} \textit{Journal of the American Statistical
  Association}, 112, 266--281.

\end{thebibliography}
\endgroup

\newpage
\renewcommand{\thesubsection}{S\arabic{subsection}}
\renewcommand{\theequation}{S\arabic{equation}}
\renewcommand{\thefigure}{S\arabic{figure}}
\renewcommand{\thetable}{S\arabic{table}}
\setcounter{equation}{0}
\setcounter{figure}{0}
\setcounter{table}{0}
\setcounter{page}{1}

\end{document}